\documentclass[prd, preprint, longbibliography, 12pt]{revtex4-1}

\usepackage{commath}
\usepackage{amsmath}
\usepackage{amssymb}
\usepackage{setspace}
\usepackage{graphicx}
\usepackage{natbib}
\usepackage{float}
\usepackage{tensor}
\usepackage[utf8]{inputenc}
\usepackage{amsfonts}
\usepackage{braket}
\usepackage{esint}
\usepackage{breqn}
\usepackage{IEEEtrantools}
\usepackage{hyperref}

\usepackage[english]{babel}
\usepackage{graphicx}
\usepackage{amsmath}

\usepackage{amssymb}
\usepackage{multirow}
\usepackage{url}
\usepackage{tikz}
\usepackage{braket}
\usetikzlibrary{decorations.markings}
\usepackage{subcaption}
\usepackage{cancel}
\usepackage{mathtools}
\usepackage[utf8]{inputenc}
\usepackage{xargs, ifthen}
\usepackage[breakwords]{truncate}

\usepackage{hyperref}
\DeclareMathAlphabet{\pazocal}{OMS}{zplm}{m}{n}

\allowdisplaybreaks

\def\s[#1\s]{\begin{align}\begin{split}#1\end{split}\end{align}}
\def\[#1\]{\begin{align}#1\end{align}}

\begin{document}
 
 %

\begin{center}
{ \large \bf Trace dynamics and division algebras: towards quantum gravity and unification }\\

\vskip 0.2 in

{\large{\bf  Tejinder P.  Singh }}

\medskip

{\it Tata Institute of Fundamental Research, Homi Bhabha Road, Mumbai 400005, India}\\


{\tt e-mail:  tpsingh@tifr.res.in\\}

\bigskip 

{\it Accepted  for publication in Zeitschrift fur Naturforschung A on October 4, 2020}

{\it v4. Submitted  to arXiv.org [hep-th] on November 9, 2020}


\end{center}

\bigskip

\centerline{\bf ABSTRACT}
\noindent  We have recently proposed a Lagrangian in trace dynamics at the Planck scale, for unification of gravitation, Yang-Mills fields, and fermions. Dynamical variables are described by odd-grade (fermionic) and even-grade (bosonic)  Grassmann matrices. Evolution takes place in Connes time. At energies much lower than Planck scale, trace dynamics reduces to quantum field theory. In the present paper we explain that the correct understanding of spin requires us to formulate the theory in 8-D octonionic space. The automorphisms of the octonion algebra, which belong to the smallest exceptional Lie group $G_2$, replace space-time diffeomorphisms and internal gauge transformations, bringing them under a common unified fold. Building on earlier work by other researchers on division algebras, we propose the Lorentz-weak unification at the Planck scale, the symmetry group being the stabiliser group of the quaternions inside the octonions. This is one of the two maximal sub-groups of $G_2$, the other one being $SU(3)$, the element preserver group of octonions. This latter group, coupled with $U(1)_{em}$, describes the electro-colour symmetry, as shown earlier by Furey. We predict a new massless spin one boson [the `Lorentz’ boson] which should be looked for in experiments. Our Lagrangian  correctly describes three fermion generations, through three copies of the group $G_2$, embedded in the exceptional Lie group $F_4$. This is the unification group for the four fundamental interactions, and it also happens to be the automorphism group of the exceptional Jordan algebra. Gravitation is shown to be an emergent classical phenomenon. Whereas at the Planck scale, there is present a quantised version of the Lorentz symmetry, mediated by the Lorentz boson. We argue that at sub-Planck scales, the self-adjoint part of the octonionic trace dynamics bears a relationship with string theory in eleven dimensions.

\bigskip

\tableofcontents

\section{Introduction and Summary}
In earlier work by Singh \cite{Singh:2012} we have argued that there ought to exist a reformulation of quantum field theory which does not depend on classical time. Considering that quantum mechanics can be thought of as a non-commutative geometry in phase space, it might be possible to obtain the desired reformulation by raising space-time points also to the status of operators, and attempt a unified description of quantum systems and non-commutative space-time by an application of Connes' non-commutative geometry programme \cite{Connes2000}. It turns out that it is possible to accomplish this, provided dynamics is described by the laws of trace dynamics \cite{Adler:04, Adler:94, AdlerMillard:1996, maithresh2019b, RMP:2012}, assumed to hold at the Planck scale. Since classical space-time is now lost, we  do not use rules of quantum field theory at the Planck scale. Instead, we employ trace dynamics, which is a matrix-valued Lagrangian dynamics for matter and gravitational degrees of freedom. Further, because we are working within the framework of non-commutative geometry, the theory inherently possesses an intrinsic notion of time, [over and above the time that was part of classical space-time] which we have named Connes time. The existence of the Connes time parameter owes itself to the Tomita-Takesaki theory and the so-called Radone-Nykodym theorem \cite{connes1994, Takesaki1, Takesaki2, Nykodym}.
For a lucid explanation by Connes see also \href{http://noncommutativegeometry.blogspot.com/2007/03/time.html}{this\ link}.
 Trace dynamics describes the evolution of matrix-valued degrees of freedom [Grassmann number-valued matrices] in Connes time, and this is assumed to be the appropriate description of dynamics at the Planck scale. The theory possesses a novel conserved charge [the Adler-Millard charge] as a result of global unitary invariance of the trace Lagrangian and trace Hamiltonian. The only fundamental constants of the theory are Planck time and Planck length, and a constant $C_0$ with dimensions of action. And associated with every degree of freedom is a length scale $L$, measured in units of Planck length. Connes time is measured in units of Planck time, and action is measured in units of $C_0$. Thus every variable in the action and Lagrangian can be expressed in a dimensionless manner, as if it were just a number: a real number, a complex number, or a Grassmann number, [or as we will introduce in the present paper, a quaternion, or an octonion, or a sedenion]. It also turns out that the Lagrangian of the theory [this being the trace of a matrix polynomial] is not necessarily a real scalar: the matrix polynomial will in general have an anti-self-adjoint part. 

Given this Planck scale Lagrangian matrix dynamics, we next ask, what does trace dynamics approximate to, if we are not observing the dynamics over Planck time resolution [in Connes time] but over times much larger than Planck time? The answer to this question can be found by applying the standard techniques of statistical thermodynamics to the phase space evolution of the matrix dynamics. This is equivalent to coarse-graining the underlying theory over time scales much larger Planck time, and asking what the approximate emergent dynamics is. It turns out that, if the anti-self-adjoint part of the underlying Lagrangian is negligible, then the emergent dynamics is the sought for quantum theory without classical time.  Provided we identify the statistically averaged matrix (equivalently operator) degrees of freedom with corresponding dynamical variables [operators] of quantum theory. The Adler-Millard charge, which has dimensions of action, gets equipartitioned over all matrix degrees of freedom, and is identified with Planck's constant $\hbar$. Quantum commutation and anti-commutation relations emerge, and the statistically averaged Hamilton's equations of motion of the underlying trace dynamics now become Heisenberg equations of motion of the emergent quantum theory without classical time. Evolution continues to be described via Connes time, as there is no classical space-time, yet. This is also a quantum theory of gravity, which we have named Spontaneous Quantum Gravity \cite{Singh:sqg, singh2019qf}.

If, in the underlying trace dynamics, sufficiently many degrees of freedom get entangled [entanglement is a property more general than quantum theory], then the anti-self-adjoint part of the operator Hamiltonian  is no longer negligible, and the evolution becomes non-unitary. This leads to a breakdown of superpositions, fermionic degrees of freedom get localised, and classical space-time, as well as classical macroscopic objects, emerge. We have shown that for a suitably chosen underlying Lagrangian, the emergent degrees of freedom obey the laws of classical general relativity coupled to matter \cite{maithresh2019}.

Those degrees of freedom in the underlying theory which are not sufficiently entangled remain non-classical. They can be described as before, by the laws of trace dynamics, with evolution in Connes time. Or, their dynamics can be described with respect to the emergent space-time already present, with evolution now described with respect to the conventional classical time [not Connes time]. These are the laws of quantum field theory. This is how conventional quantum field theory is recovered from an underlying formulation which does not depend on classical time. The underlying theory also leads to a promising new quantum theory of gravity, i.e. the afore-mentioned spontaneous quantum gravity.

In trace dynamics,  the matrix degrees of freedom are described by Grassmann-number valued matrices [over the field of complex numbers]. Odd-grade Grassmann matrices $q_F$ are called fermionic and they describe fermions. Even grade Grassmann matrices $q_B$ are called bosonic and describe bosons. However, in trace dynamics, neither bosons nor fermions have a constant spin, even though spin angular momentum can be defined \cite{Singhspin} [recall that there is no $\hbar$ in trace dynamics, so there cannot be a fixed integer or half-integer spin: discrete spin is an emergent property of bosons and fermions. To start with their definition is only whether they are described by even-grade or odd-grade Grassmann matrices].

Inspired by a result  \cite{Chams:1997} from Riemannian geometry and its significance for non-commutative geometry, we have introduced in trace dynamics the important concept of an `atom of space-time-matter' also called an `aikyon' \cite{MPSingh}. The said result from geometry is that given a Riemannian spin manifold, and the conventional Dirac operator $D_B$  on this manifold, the curvature of the manifold is related to the Dirac operator. More precisely, the trace $Tr [D_B]^2$, is proportional to $\int  d^4 x\; \sqrt{g}\; R$ where $R$ is the Ricci scalar, in a truncated heat kernel expansion of the trace in powers of a parameter with dimensions of area [in our case this will be $L_P^2$]. This relation between this trace and curvature is called the spectral action principle, because the action of general relativity is being expressed in terms of spectrum of the Dirac operator. Now if one no longer has a Riemannian manifold, i.e. if the manifold is replaced by a non-commutative geometry, one still has the Dirac operator in the new geometry, and this can be gainfully used to continue to talk of concepts such as curvature, and the metric, even though the geometry is no longer commutative.

In trace dynamics, the Lagrangian is the trace of a matrix polynomial, which is then integrated over time [or over space-time volume element in the continuum limit] to arrive at the action. In Adler's original version of trace dynamics, space-time is assumed to be Minkowski flat, even though the theory is operating at the Planck scale. Clearly, this is meant only as an approximation, it being understood that further development of the theory will incorporate gravitation as well. The fact that in Riemannian geometry and in its generalisation to non-commutative geometry, the Einstein-Hilbert action can be expressed as a trace, is a clue as to how gravity should be included in trace dynamics. We introduce a self-adjoint bosonic Grassmann matrix $q_B$ in trace dynamics, such that 
\begin{equation}
D_B \equiv \frac{1}{Lc} \frac{dq_B}{d\tau}
\end{equation}
where $L$ is a length scale associated with the `atom' of space-time $q_B$. The contribution of gravity to the trace Lagrangian of trace dynamics  will be assumed to be proportional to 
$\sim Tr [\dot{q}_B^2]$ where a dot denotes derivative with respect to Connes time $\tau$. The trace dynamics action for gravity will be $S \sim \int d\tau \ Tr [\dot{q}_B^2]$. Note that what was the spectral {\it action} in Riemannian geometry has now become the spectral {\it Lagrangian} in trace dynamics! In the classical limit where space-time emerges, this action goes to $S\sim \int d\tau\; \int d^4x \; \sqrt {g} R$. This is a clear indicator that there are two times in the theory, the absolute universal Connes time, and the conventional time that is part of a Lorentz-invariant 4-D space-time. We emphasise that ours is a Lorentz-invariant theory at the Planck scale. Connes time is an additional time parameter and its existence must not be interpreted as violation of Lorentz invariance. In fact, as we will newly show in this paper, the `atom' $\dot{q}_B$ of space-time is the `Lorentz gauge field' whose quantisation defines a new particle, a massless spin one Lorentz boson. $\dot{q}_B$ does not directly describe gravity. There is in fact no gravitation at the Planck scale. There is only the Lorentz gauge field. Gravitation emerges only in the classical limit as a condensate of a large number of such `atoms' of space-time [gravitation is an emergent phenomenon \cite{jacobson_1995, paddy}]. 

Next, we would like to have an action for fermions in our theory, on the same footing as the action for an `atom' of space-time. Thus, unlike the earlier non-commutative geometry based approaches to the standard model \cite{Schucker2000spin}, where the spectral action including fermions takes the form $\{(Tr [D_B]^2) + {\rm Fermionic\ Action}\}$, we would like to have a trace dynamics action of the form 
$Tr [D_B + D_F]^2$,  where $D_F$ is a newly introduced (non-self-adjoint) operator which represents fermions.  In the emergent classical limit, this [symbolically] takes the form
$(D_B^2 + D_B D_F + D_F^2)$ where the first term becomes the Einstein-Hilbert action, the second term becomes the Dirac action for a fermion (leading to the Dirac equation), and the last term the Higgs boson, which essentially comes for free in this construction.

Bringing fermions inside the square is challenging, but it can be done \cite{maithresh2019}. We define an odd-grade (hence fermionic) Grassmann matrix $\dot{q}_F$ to represent fermions, and define the `fermionic Dirac operator' $D_F\equiv (1/Lc) \; d{q}_F / d\tau$. We define an `atom' of space-time-matter, i.e. an aikyon, as $q\equiv q_B + q_F$. This is nothing but the splitting of a general Grassmann matrix into its even-grade (i.e. bosonic) and odd-grade (i.e. fermionic) part. One can construct an action principle for the aikyon in trace dynamics provided one gives up the squared Dirac operator in favour of a bilinear form! We introduce two constant fermionic (odd) Grassmann numbers $\beta_1$ and $\beta_2$ which must not be equal to each other. The action principle for an aikyon takes the form
\begin{equation}
S \sim \int d\tau\; Tr \bigg\{ \frac{L_P^2}{L^2}[\dot{q}_B + \beta_1 \dot{q}_F^\dagger  ] \times [\dot{q}_B + \beta_2 \dot{q}_F] \bigg\}
\end{equation} 
It is not possible to make a consistent theory if $\beta_1 = \beta_2$. This immediately leads us to think of the aikyon as a 2-D object  evolving in Connes time. In all likelihood, the aikyon is the same object as the closed string of string theory. However, at the Planck scale the dynamics of the aikyon is described by the laws of trace dynamics, not by the laws of quantum field theory. Hence our theory is different from string theory, even though the aikyon is likely the closed string. We would like to suggest that string dynamics should be trace dynamics based on a (non-self-adjoint) Lagrangian. This trace (matrix)  dynamics is not to be quantised. Rather, quantum (field) theory is emergent from it. 

The transition from classical general relativity [gravitation coupled to relativistic point particles] to a pre-quantum, pre-space-time matrix dynamics is the most crucial step in the development of the Aikyon Theory. Hence we elaborate on this transition, for the sake of clarity. The philosophy of trace dynamics is to start from a classical Lagrangian dynamics, and raise the c-number configuration and momentum variables to the status of matrices. These matrices are defined such that their eigenvalues are the original c-number configuration and momentum variables themselves. The new Lagrangian is defined as the trace of the matrix polynomial which results when the configuration variables and velocities in the original Lagrangian are replaced by matrices. The resulting Lagrangian matrix dynamics is trace dynamics - a pre-quantum dynamics - from which quantum theory is emergent. The trace Lagrangian must have an additional feature arising naturally: the corresponding Hamiltonian must not be self-adjoint in general. When the anti-self-adjoint part is small and ignorable, the emergent low-energy theory is quantum [field] theory. Under suitable circumstances [adequate entanglement amongst degrees of freedom] the anti-self-adjoint part becomes significant, and spontaneous localisation results. In this process, the trace Lagrangian is mapped to one of its eigenvalues, thereby converting the trace to the original c-number self-adjoint Lagrangian. The role of the anti-self-adjoint part is to provide imaginary stochastic fluctuations which, in the nature of objective collapse models, enable the quantum-to-classical transition. In this way, quantum theory, as well as classical dynamics, are both recovered from the pre-quantum trace dynamics, as suitable low energy emergent approximations. 

In trace dynamics, space-time is Minkowski flat. We would like to incorporate gravity, by raising classical space-time and its geometry to matrix status. For this we must first identify a suitable way to express the classical theory. Let us begin by schematically writing down the action for classical general relativity:
\begin{equation}
S_{GR} \sim \int d^4x \; \sqrt{g} \;  \frac{R}{L_P^2} + \sum_i m_i \int ds
\end{equation}
[It is worth noting the remarkable fact that, in spite of being second order, the Einstein equations are linear in the source mass, and not quadratic, unlike the Klein-Gordon equation. This is an indicator that the mass term arises as a cross-term when a square is opened up.] We now employ the Dirac operator on this classical manifold, to express the gravity part of the action in terms of the eigenvalues $\lambda_i$ of the Dirac operator:
\begin{equation}
Tr \; [L_P^2 \; D_B^2] \sim \int d^4x \sqrt{g} \; \frac{R}{L_P^2} + {\cal O}(L_P^0)\sim L_P^2 \sum_i \lambda_i^2
\label{grraction}
\end{equation}
so that the full action is
\begin{equation}
S_{GR} \sim L_P^2 \sum_i \lambda_i^2 \ + \ \sum_i m_i \int d^4x \sqrt{g} \delta({\bf x} -{\bf x_0}(t))
\label{ontheway}
\end{equation}
To transit to the pre-space-time theory, {\it each} of the eigenvalues $\lambda_i$  is now raised to the status of a matrix $D_{Bi}$, this being the original Dirac operator itself, so that we now have as many copies of the Dirac operator as the number of its eigenvalues, and
\begin{equation}
\lambda_i \rightarrow D_{Bi} \equiv \frac{1}{L}\frac{dq_{Bi}}{d\tau}
\end{equation}
Here, $q_{Bi}$ is the configuration variable, now a matrix/operator, which defines the  $i$-th atom of space-time.  Connes time $\tau$ has been brought in, because we are now in the domain of non-commutative geometry: the matrices $q_{Bi}$ do not commute with each other. Classical space-time points have been lost, and we have a pre-quantum, pre-space-time dynamics. We refer to this as generalised trace dynamics [GTD]. The action for the $i$-th space-time atom will be $\int d\tau\; Tr [D_{Bi}^2]$, and the GTD action for the gravitation part of the theory will hence be
\begin{equation}
S_{GTD} \sim L_P^2 \sum_i \int d\tau \; Tr [D_{Bi}^2]
\end{equation} 
These are the atoms of space-time, from which space-time is emergent, after critical entanglement leads to spontaneous localisation. However, in our theory, matter in the form of fermions [coupled to gravity] is absolutely essential for the emergence of space-time [no matter, no space-time]. The points of the space-time manifold are defined by the c-number positions that fermions acquire after spontaneous localisation; these positions being the eigenvalues of the fermionic Dirac operator $D_F\equiv (1/Lc) \; d{q}_F / d\tau$ introduced 
above.  We can define this fermionic Dirac operator more precisely, by first {\it assuming} that there are as many fermions as the number of eigenvalues of the Dirac operator $D_{B}$, and then by choosing the $i$-th operator $\dot{q}_{Fi}$ such that the eigenvalues of $\dot{q}_{Bi}\dot{q}_{Fi}$ [upon inspection of the matter action] are proportional to $m_i\delta (\bf{x}-{\bf x_{0}}(t))$. Thus, the GTD action for gravity and fermions can now be schematically written as
\begin{equation}
S_{GTD} \sim \sum_i S_i \sim \sum_i \int d\tau \; \bigg(Tr [\dot{q}_{Bi}^2] + Tr [\dot{q}_{Bi} \dot{q}_{Fi}]\bigg) 
\label{accgtd}
\end{equation}
This action paves the way for introducing the aikyon - an atom of space-time-matter. Spontaneous localisation sends this trace Lagrangian  above to the one shown in Eqn. (\ref{ontheway}) [each trace goes to an eigenvalue, and the Connes time integral stays as such]. Then from (\ref{ontheway}) we can re-construct the general relativity action (\ref{grraction}). We emphasise that the action (\ref{accgtd}) is henceforth to be taken as the first-principles action, and no longer dependent on the general relativity action. All we need  to know is that the eigenvalues are those of the Dirac operator $D_B$ and of the fermionic Dirac operator $D_F$, which are in turn related to $q_{Bi}$ and $q_{Fi}$. Space-time arises from `collapse of the wave-function' \cite{Singh:2019, maithresh2019}. 

Taking clue from this form of the action above, we propose the following fundamental form for the action of an aikyon, thereby bringing the fermions `inside the square':
\begin{equation}
 S\sim \int d\tau\; Tr \bigg\{ \frac{L_P^2}{L^2}[\dot{q}_B + \beta_1 \dot{q}_F^\dagger  ] \times [\dot{q}_B + \beta_2 \dot{q}_F] \bigg\}
\end{equation}
This is the action for the $i$-th aikyon, and the total action for the universe is the sum over all aikyons, of such individual terms. For most part of this paper, we will work with the action of only one aikyon. It is implicit that the total number of space-time dimensions is $4+1=3+1+1$, with the fifth dimension being Connes time. Soon however, we will conclude that the physical space must be doubled to have eight dimensions, making this a $8+1$-d theory. With this doubling of spatial dimensions, the action will be found to describe not just gravity, but also the weak interaction! If we think of the aikyon itself as a 2-d object [because of the two constants $\beta_1$ and $\beta_2$] then we have a 8+1+2 = 11-d theory.

In summary, we have followed in the footsteps of trace dynamics to generalise the pre-quantum theory to a pre-quantum pre-space-time theory. From this theory, quantum theory and classical gravitation are emergent as low-energy approximations.

This Lagrangian thus defined can be extended to include Yang-Mills fields $q_B$ \cite{MPSingh} via the following kind of construction
\begin{equation}
S \sim \int d\tau\; Tr \bigg\{ \frac{L_P^2}{L^2}\bigg[(\dot{q}_B + \beta_1 \dot{q}_F^\dagger) + i\frac{\alpha}{L} (q_B^\dagger + \beta_1 q_F^\dagger)\bigg] \times \bigg[ (\dot{q}_B + \beta_2 \dot{q}_F) + i\frac{\alpha}{L} (q_B +\beta_2 q_F)\bigg] \bigg\}
\label{ymi}
\end{equation} 
Here, $q_B$ describes Yang-Mills fields, $q_F$ as well the earlier $\dot{q}_F$ describe fermions, and $\dot{q}_B$ continues to be self-adjoint. $\alpha$ is the dimensionless coupling constant which describes the coupling of Yang-Mills fields to fermions. This is the form of the Lagrangian for an aikyon (to be made precise in the next section). It is a unified description of gravitation, Yang-Mills fields, and fermions, and is hence a candidate for investigating the unification of gravitation with the standard model of particle physics. This approach to unification will be studied in the present article.

At the Planck scale, the universe is made of enormously many aikyons; each aikyon being an elementary particle (fermion) plus the fields (bosonic) it produces. No distinction is made between the source and the `produced' field. This makes sense; if there is no space-time on which the field can live, far away from the source, nor a space-time in which the source is embedded, it is only reasonable to think of $q_B$ and $q_F$ as parts of the same entity, namely the aikyon $q=q_B + q_F$. The following features that we have introduced in trace dynamics, play a very important role in unification: Connes time, the Dirac operator as the atom of space-time  $\dot{q}_B$, the aikyon  concept, and the identification of $\dot{q}_B$ with the Lorentz interaction, and $q_B$ as Yang-Mills interaction [added with the imaginary $i$ as a factor]; $q_B$ and $\dot{q}_B$ have different interpretations.

Aikyons interact via entanglement, and sufficient entanglement leads to spontaneous localisation of the fermionic part, giving rise to emergent classical space-time and gravity as a condensate, classical gauge-fields, and classical material point sources, obeying the rules of classical general relativity.

In a recent paper \cite{Singhspin} we have shown, in the following manner, that quantum spin is defined as the canonical angular momentum corresponding to the time variation of a certain angle. We have seen above in the Lagrangian (\ref{ymi}) that the Yang-Mills field is introduced with an $i$ factor (only then does one get finite solutions to the equations of motion \cite{MPSingh}). This permits us to think of $\dot{q}_B$ and $q_B$ as if they are the real and `imaginary' components of $\dot{q}_B + i q_B$.  Writing this sum in the equivalent form $R_B \exp i\theta_B$ and substituting it in the Lagrangian permits us to define spin angular momentum as proportional to $\dot{\theta}_B$; spin results from the time rate of change of the phase that relates the Lorentz field with Yang-Mills field. It is shown that this definition gives the correct conventional interpretation of spin. Also bosonic Grassmann matrices are shown to have integral spin, and fermionic matrices are shown to have half-integral spin, in the emergent theory. Instead of deriving statistics from spin, we have derived spin from statistics, thus providing a simple proof of the spin-statistics connection. 

In constructing this definition of spin, we noticed something very curious. $\dot{q}_B$ lies along four real directions [space-time]. It is as if $q_B$ does not lie in space-time, but along an orthogonal set of four imaginary directions (it is natural to expect four components of $q_B$, if $\dot{q}_B$ has four). That is a total of eight directions. This has far-reaching implications! 
Firstly, it looks like the right place for this Lagrangian is phase space, not space-time: we have `position' $q_B$ and velocity $\dot{q}_B$. Both have their own independent interpretation: just like the $(q, p)$ pair in phase space. Secondly, ours is a non-commutative space, and eight directions immediately suggests octonions! There is a rich history of relating octonions to the standard model, and those findings  \cite{Dixon, Gursey, f1, f2, f3, Chisholm, Trayling, Dubois_Violette_2016, Todorov:2019hlc, Dubois-Violette:2018wgs, Todorov:2018yvi, Todorov:2020zae, ablamoowicz, baez2001octonions, Baez_2011,  baez2009algebra}    open the gateway for relating our Lagrangian  to division algebras, allowing us to unify the Lorentz symmetry with the standard model, and propose a unified theory in the present paper. Thirdly, it is known that eleven dimensional string theory [M-theory] is akin to having ten space-time dimensions, plus the eleventh dimension possibly serving as the second time [Connes time ?]. It is also known those ten space-time dimensions are equivalent to eight octonionic dimensions \cite{Baez}. So it could well be that the self-adjoint part of the Hamiltonian of our theory, at energies below Planck scale, describes the same theory as string theory. However, we do not need dimensional compactification or Calabi-Yau manifolds. Dimensional reduction is naturally achieved for classical systems via the dynamical process of spontaneous localisation. Quantum systems continue to remain and evolve in eight dimensional octonionic space. An aikyon can be thought of as evolving in this 8-D non-commutative phase space in Connes time. The algebra automorphisms of the octonions take the place of gauge transformations [internal symmetries]. But they also take the place of space-time diffeomorphisms. There is no longer a Diff M at the Planck scale. Instead, the algebra automorphisms take the place of Diff M and of internal gauge transformations, and unify them into one concept. General covariance and gauge invariance are unified, and the group formed by the automorphisms of octonions is the smallest of the exceptional Lie groups, $G_2$. This then will be the symmetry group for one generation of fermions unified with the gauge interactions and with Lorentz symmetry.

In our theory, $\dot{q}_B$ will have four octonionic directional components, including the real direction, so that ${\dot q}_B$ will be the quaternion part of the octonion. $q_B$ will occupy the other four octonion directions. Together, the $q_B$ and $\dot{q}_B$ will have eight directions and components - they are a set of eight complex-numbered bosonic Grassmann matrices over eight octonion directions. Similarly, the $\dot{q}_F$ and $q_F$, the fermions, together have eight octonion components. Our quadratic Lagrangian has the structure of complex octonions acting on themselves; hence the relevance of division algebra studies for our theory.

With this background of earlier work, we are now ready to relate the Lagrangian of the theory to the standard model, and to division algebras, sedenions, and the exceptional Jordan algebra. Most importantly, in our theory we were in need of a physical non-commutative space in which the aikyon lives, and we have found one in the octonions. On the other hand, the profound studies relating division algebras to the standard model are badly in need of a Lagrangian! Their story is incomplete if there are only symmetries but no Lagrangian whose symmetries those are. The Lagrangian we have constructed in trace dynamics turns out to be perfect for relating to division algebras. There is no room for manoeuvre, no fine-tuning and no free parameters at all. If the predictions of this Lagrangian disagree with known physics [our theory is eminently falsifiable] then this theory will be totally ruled out. It cannot be saved by making alterations or adjustments. Division algebras cannot be applied to the conventional standard model Lagrangian, because that Lagrangian resides in space-time and is generally covariant. It is not in need of algebra automorphisms. These algebra automorphisms become key at the Planck scale. And even more importantly, the trace dynamics Lagrangian does not have to be quantised. It already describes a dynamics from which quantum theory is emergent. And the trace Lagrangian is invariant under global unitary transformations constructed from the generators of these automorphisms.

The relevance of algebras to particle physics goes back to the 1934 paper by Jordan, von Neumann and Wigner \cite{Jordan}, and Albert \cite{Albert1933} who discovered the significance of the exceptional Jordan algebra: the algebra of $3\times 3$ Hermitean matrices with octonionic entries. There was then some lull until in the 1970s when Gunaydin and Gursey \cite{Gunaydin2} showed that important properties of quarks and leptons can be inferred from the algebra of octonions acting onto themselves. This was followed by important studies by various researchers including Dixon and Baez. From our point of view, the fast-paced developments during the last five years or so are extremely significant, and are already pointing to the correct symmetry group for unification. In arriving at the findings of the present paper, we have been inspired by the recent research of Furey \cite{f1}, of Stoica \cite{Stoica}, of Gillard and Gresnigt \cite{Gillard_2019}, and of Dubois-Violette and Todorov \cite{Dubois-Violette:2018wgs}. In arriving at the unified theory proposed in this paper, we build heavily on their work, and explain it in the context of our Lagrangian, and show how perfectly our Lagrangian fits several of their findings. 

In her 2016 Ph. D. thesis \cite{f1}, Furey explains, building on the work of earlier researchers,  the significance of minimal left ideals constructed from the algebra of complex quaternions acting on themselves. The complex quaternions give a faithful representation of the Clifford algebra $C\ell(2)$. Spinors are minimal left ideals of Clifford algebras.  The left and right handed Weyl spinors are minimal left ideals made from the action of $C\ell(2)$ generators on the idempotent. These Weyl spinors transform correctly under $SL(2, C)$, which is the Lorentz group and is also the group of automorphisms of the complex quaternions. In our Lagrangian, the symmetry group leaving the first term invariant is this automorphism group. Hence, as per the aikyon concept, this term gets identified with the Lorentz `interaction', which is mediated by the Lorentz boson.

In an analogous manner Furey then shows that the complex octonions can be used to deduce the Clifford algebra $C\ell(6)$ which has six generators. These generate an eight dimensional basis. Now, it turns out that the generators have a $U(3)\sim SU(3) \times U(1)/{\mathbb Z}_3$ symmetry. The $SU(3)$ is an element preserving subgroup of $G_2$, the automorphism group of the octonions, and the $U(1)$ is a number operator made from these generators. Minimal left ideals are made by left multiplying $C\ell(6)$ on the idempotent, giving rise to the said 8-D basis. Now, the $U(3)$ symmetry imposes a definite discrete structure on this basis, compelling us to identify it with one generation of quarks and leptons. Thus the algebra of complex octonions describes the eight fermions of one generation in the standard model, which is beautiful. The unbroken $SU(3)_c\times U(1)_{em}$ symmetry of the standard model can be related to a division algebra. Three of the eight terms in our Lagrangian relate to this $U(3)$ symmetry.

But what about the $SU(2)$ weak symmetry, and what about the unbroken electro-weak symmetry? How to relate them to a division algebra? In a 2018 paper \cite{f3} Furey employs four of the electro-colour generators to make an $SU(2)$ symmetry from the Clifford algebra $C\ell(4)$. This symmetry acts correctly on the leptons, as expected from the standard model. However, it would appear that this cannot be a fundamental construct, because the generators have come from the electro-colour algebra. And the octonions can give only one representation of $C\ell(6)$, which is already used up. So the complex octonions seem like an unlikely candidate for explaining the weak symmetry. In another elegant 2018 paper, Stoica showed \cite{Stoica}  that two copies of $C\ell(6)$, which he proposes to identify, describe correctly the symmetries of one generation of the eight quarks and leptons, {\it including} the Lorentz symmetry. His construction is not concerned with division algebras. Again, four of the six generators of the weak-Lorentz sector can in principle be constructed from the electro-colour sector. This is how things stood until our present paper.

A careful analysis of these two papers, which we describe in detail in the next section, leads us to the propose the Lorentz-weak unification, which we also call the gravito-weak symmetry. In order to arrive at this conclusion, we have to look carefully at the two maximal sub-groups of $G_2$ \cite{Yokota}. One of them is the element preserving group of the octonions; it is the $SU(3)$. The other maximal sub-group is the stabiliser group of the quaternions in the octonions. It is $SU(2)\times SU(2)/{\mathbb Z}_2$, and does not seem to have gotten the attention it deserves. Although Todorov and Drenska \cite{Todorov:2018yvi} do prove its existence. It has a sub-group $SO(3)$ which is the group of automorphisms of the quaternions. The group extension of this $SO(3)$ is an $SU(2)$ which happens to be the element preserving group of the quaternions. Moreover, the two maximal subgroups have a $U(2)$ intersection, of which the said $SU(2)$ is a normal subgroup. Todorov and Dubois-Violette \cite{tkey}  note that this $U(2)$ is precisely the Weinberg-Salam electro-weak model! Taking clue from this group intersection, we propose the gravito-weak symmetry: At the Planck scale, the Lorentz symmetry is unified with the weak symmetry, and hence with electro-weak. The  symmetry group for the gravito-weak symmetry is the stabiliser group of the quaternions inside the octonions. When spontaneous localisation separates spacetime and hence the Lorentz symmetry, the electro-weak becomes part of the internal symmetries, along with $SU(3)_c$. Three terms in our Lagrangian perfectly describe the gravito-weak symmetry and lead us to predict the Lorentz boson. Another two terms describe the bosons, and the remaining two describe what are likely the Higgs bosons. Hence the Lagrangian perfectly describes the standard model bosons, and  one fermion generation. The $U(2)$ intersection of the two sub-groups, wherein the weak symmetry lies, explains why the generators of the weak symmetry can be constructed from those of $SU(3)_c$. And, to describe the Lorentz-weak symmetry, we must indeed look beyond the complex octonions and beyond  division algebras. 

Consequently, this analysis quickly leads to a proposal for unification of all the interactions. It turns out that if we retain in the Lagrangian the total time derivative terms which were dropped to arrive at the above-mentioned one generation Lagrangian, something remarkable happens. The Lagrangian now has the quadratic form of complex sedenions acting on themselves. Following the recent work of Gillard and Gresnigt \cite{Gillard_2019}, we show that the Lagrangian describes the symmetries of three fermion generations, including the unification with the Lorentz interaction. This then is a unification of all the four interactions of the known elementary particles. The symmetry group is made of three [intersecting] copies of $G_2$, which are embedded in the exceptional Lie group $F_4$. This then is the unification group of the unified theory. The self-adjoint part of our three-generation Lagrangian is related to the exceptional Jordan algebra $J_3({\mathbb O})$, whose automorphism group is again the same $F_4$. The (cubic) characteristic equation of this algebra very likely determines the masses of three generations of elementary particles, as well as values of the standard model parameters. 

The next section describes in detail our discovery of the connection between trace dynamics, division algebras, sedenions, the exceptional Jordan algebra, and a promising proposal for unification of the four known forces. Our theory is falsifiable and testable with current technology, and we also mention the predictions of our theory.

\section{Trace dynamics, division algebras, and the standard model}
In our recent paper \cite{MPSingh} we have proposed the following trace dynamics Lagrangian and action for the unification of gravity, Yang-Mills fields, and fermions  at the Planck scale [Eqn. (45) of the said paper]:
\begin{equation}
\begin{split}
 \frac{S}{C_0} = \frac{1}{2}\int \frac{d\tau}{\tau_{Pl}}\; Tr \bigg[\biggr.\frac{L_P^2}{L^2}\bigg\{\biggr.  \bigg(\dot{q}_B^2 +\frac{L_P^2}{L^2} \dot{q}_B\beta_2 \dot{q}_F + \frac{L_P^2}{L^2} \beta_1 \dot{q}_F \dot{q}_B   + \frac{L_P^4}{L^4} \beta_1\dot{q}_F \beta_2\dot{q}_F \bigg) \\- \frac{\alpha^2}{L^2} \bigg(q_B^2     
+ \frac{ L_P^2}{L^2} q_B \beta_2 q_F + \frac{ L_P^2}{L^2} \beta_1 q_F q_B +  \frac{L_P^4}{L^4} \beta_1 q_F \beta_2 q_F  \bigg)
 \biggl. \bigg\} \biggl.\bigg] 
 \end{split}
 \label{fulllag}
 \end{equation}
 This Lagrangian with eight terms arises from opening up the form shown in (\ref{ymi}) \cite{MPSingh}. That gives rise to sixteen terms, eight of which are total time derivatives and have been discarded. We will return to these discarded terms later in the paper.
 Here, $q_B$ and $q_F$ are respectively even-grade (bosonic)  and odd-grade (fermionic) Grassmann matrices,  and $L$ is a length scale associated with the aikyon, to be measured in units of Planck length $L_P$.  $C_0$ is a real constant with dimensions of action. Evolution takes place in Connes time $\tau$, and a dot denotes time derivative with respect to $\tau$. There is no classical space-time in the theory. Rather, evolution takes place in a non-commutative octonionic space, which we introduce below: this will lead us to division algebras and their deep connection with the standard model, and its unification with gravity. From this dynamics, 4-D classical space-time emerges at low energies, as a consequence of spontaneous localisation.

 In the present paper, we will work with a somewhat modified Lagrangian, for reasons that are explained in the next section. We will assume $\beta_1$ and $\beta_2$ to be two (distinct) odd-grade Grassmann elements, unlike before, when they were assumed to be constant fermionic matrices. Furthermore, we will replace $\beta_1  q_F$ by $\beta_1 {q}^\dagger_F$, and $\beta_1 \dot{q}_F$ will be replaced by $\beta_1 \dot{q}^{\dagger}_F$. The fermionic Grassmann matrix $q_F$ and its adjoint $q_F^\dagger$ will be treated as independent degrees of freedom. Thus, the Lagrangian that we will work with, in this paper, will be
 \begin{equation}
\begin{split}
 \frac{S}{C_0} = \frac{1}{2}\int \frac{d\tau}{\tau_{Pl}}\; Tr \bigg[\biggr.\frac{L_P^2}{L^2}\bigg\{\biggr.  \bigg(\dot{q}_B^2 +\frac{L_P^2}{L^2} \dot{q}_B\beta_2 \dot{q}_F + \frac{L_P^2}{L^2} \beta_1 \dot{q}^\dagger_F \dot{q}_B   + \frac{L_P^4}{L^4} \beta_1\dot{q}^\dagger_F \beta_2\dot{q}_F\bigg)  \\- \frac{\alpha^2 }{L^2} \bigg(q_B^2     
+ \frac{ L_P^2}{L^2} q_B \beta_2 q_F + \frac{ L_P^2}{L^2} \beta_1 q^\dagger_F q_B +  \frac{L_P^4}{L^4} \beta_1 q^\dagger_F \beta_2 q_F  \bigg)
 \biggl. \bigg\} \biggl.\bigg] 
 \end{split}
 \label{fulllag2}
 \end{equation}
We assume $q_B$ to be self-adjoint. However, this Lagrangian is not self-adjoint [because $\beta_1$ and $\beta_2$ are unequal] and that plays a crucial role when we use division algebras to relate to the standard model. The presence of the anti-self-adjoint part is also key for inducing spontaneous localisation and emergence of classical limit.
 
 By defining the dynamical variables $q^\dagger_1 = q_B + \beta_1 q^\dagger_F$ and $q_2 = q_B + \beta_2 q_F$ this trace Lagrangian can be elegantly written as 
 \cite{q1q2uni}
\begin{equation}
Tr {\cal L} = \frac{1}{2} a_1 a_0\;   Tr \left[\dot{q}_1^\dagger  \dot{q}_2  - \frac{\alpha^2 c^2}{L^2} q_1^\dagger  q_2  \right]
\label{lagnew}
\end{equation}
 where $S\equiv \int d\tau \; {Tr \cal L}$ and   $a_0 \equiv L_P^2 / L^2$ and $a_1 \equiv C_0 /cL_P $. This is the fundamental Lagrangian which describes an aikyon $q=q_B + q_F$ in terms of $q_1$ and $q_2$. The equations of motion, which will be discussed in the next section, follow from variation of this Lagrangian [trace derivatives] with respect to the matrix degrees of freedom $q_1$ and $q_2$. These are the defining dynamical equations of the theory, which is {\it not} quantised. Rather, quantum field theory is shown to emerge as a statistical thermodynamics approximation to the underlying trace dynamics, the latter assumed to be operating at the Planck scale. The idea being that if we are not observing the Planckian dynamics, but only observing the dynamics at low energies, we coarse-grain the underlying theory over length / time  scales much larger than Planck length / Planck time, and ask what the laws of the emergent dynamics are. These laws are those of quantum field theory. However, the Planck scale laws are those of trace dynamics, different from, and more general than,  those of quantum field theory \cite{Adler:04}. 
 

For ease of further reference, we will label the eight terms in the Lagrangian (\ref{fulllag2}), starting from the left, as terms $T_1, T_2, T_3,..,T_8$. As we will see, when we describe this Lagrangian in octonionic space, the first three terms $T_1$, $T_2$ and $T_3$ describe Lorentz symmetry of 4-D non-commutative space-time, and weak isospin, and their interaction with eight fermions of one generation, and the $SU(2)$ symmetry of weak interactions. These behave like kinetic energy terms. The  terms $T_5, T_6, T_7$, the one with the coupling constant $\alpha$, describe strong interactions and electromagnetism, the eight gluons and the photon and  their interactions with fermions, and  $SU(3)_c\times U(1)_{em}$ symmetry. These are like potential energy terms. The terms $T_4$ and $T_8$ form the heart of the division algebra program, from our point of view, and explain how minimal left ideals made from action of complex octonions onto themselves determine properties of quarks and leptons. These terms  could also possibly describe the  Higgs bosons, as composites of fermions. We can also pair the terms as $(T_1, T_5)$, $(T_2, T_6)$, $(T_3, T_7)$ and then they can be thought as the [kinetic energy + potential energy] of three different `particles': i.e. $(q_B, q_F, q_F^\dagger)$. Similarly in (\ref{lagnew}) the potential energy terms describe the strong and electromagnetic interactions of the fermions, and the kinetic energy terms describe their gravitational and weak interactions. 
 
 
 \subsection{An octonionic space for the Lagrangian}
 In our recent papers by Singh et al. \cite{Singhspin, q1q2uni} we have motivated why the space of quaternions and octonions is the appropriate setting for describing the dynamics of an aikyon using the above Lagrangian, and for relating this dynamics to the standard model. We now develop this construction in full detail. The aikyon evolves in an octonionic coordinate system in Connes time, and the various bosons and fermions of the standard model relate to different directions in this 8-D coordinate system. The Lagrangian above describes their dynamics and interactions.
 
 An octonion has one real direction, which we denote as $e_0$, and seven imaginary directions, $(e_1, e_2, ..., e_7)$. The square of the real direction is unity: $e_0^2=1$, and the square of each imaginary direction is $-1$. We write the octonionic coordinate system as $(e_0, e_1, e_2, e_3, e_4, e_5, e_6, e_7)$.
 A quaternion has one real direction $e_0$ and three imaginary directions which we will label $(e_1, e_2, e_4)$. The quaternionic product rule is 
 \begin{equation}
 e_1 \times e_2 = - e_2 \times  e_1 = e_4; \qquad e_2 \times e_4 = - e_4 \times e_2 = e_1, \qquad e_4 \times e_1 = - e_1 \times e_4 = e_2
 \end{equation}
 We will work with complex quaternions and complex octonions. The product rule for the octonionic directions is given by the Fano plane. For our ready reference, we display these products explicitly below, in Table I. The seven imaginary directions come in quaternionic triples, meaning that every triple obeys the quaternionic product rule, and along with unity forms a closed quaternionic sub-algebra of the octonion algebra. In our notation, the seven triples are: $(e_1, e_2, e_4), (e_3, e_4, e_6), (e_6, e_1, e_5), (e_5, e_2, e_3), (e_3, e_7, e_1), (e_5, e_7, e_4), (e_6, e_7, e_2)$.
 \begin{figure}[!htb]
        \center{\includegraphics[width=\textwidth]
        {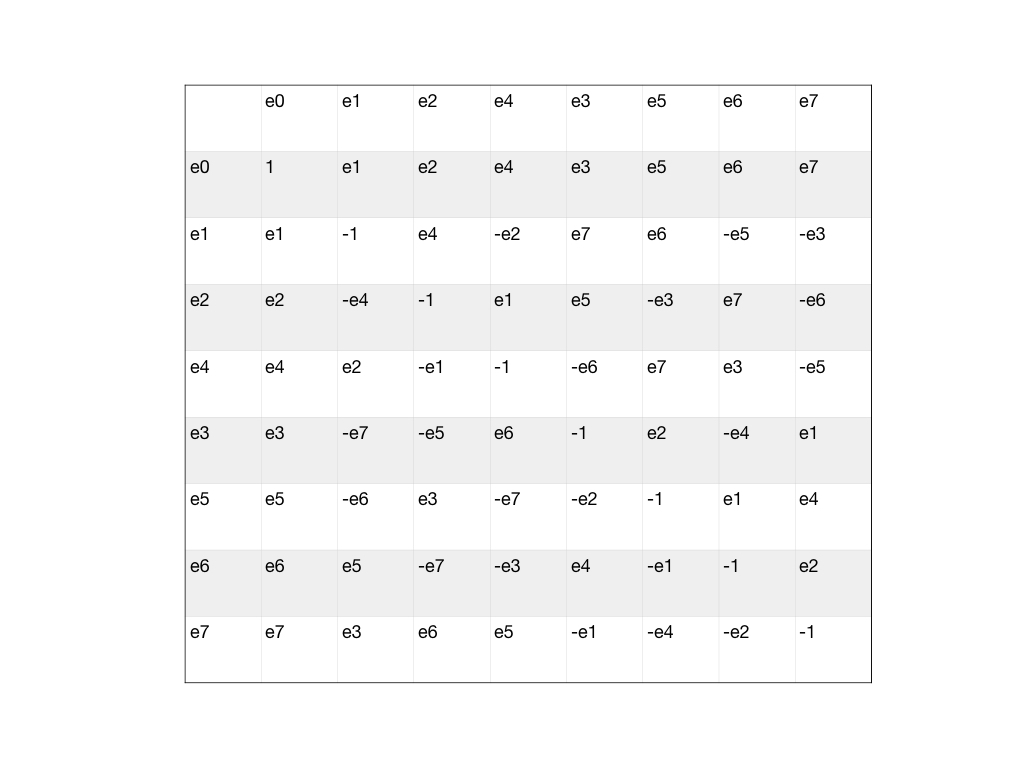}}
        \caption{\label{fig:my-label} The multiplication table for two octonions. Elements in the first column on the left, left multiply elements in the top row. We follow the notation of \cite{f1}. $(e_0, e_1, e_2, e_4)$ form a quaternion that emerges as space-time.}
      \end{figure}
 The 8x8  product table  is quite remarkable, and deserves closer attention. We think of the table as made of 4x4 sub-tables, to which we give the self-explanatory names Top-left, Top-right, Bottom-left, Bottom-right. These four sub-tables have interesting properties. The Top-left forms the quaternionic sub-algebra of octonions which has the real direction; it is the only sub-algebra in the algebra of octonions which has the real direction in it. It is made from the directions $(e_0, e_1, e_2, e_4)$ and we assign the bosonic Grassmann matrix $\dot{q}_B$, which describes gravity and weak interactions in the above Lagrangian,   these four directions.  We are assuming that the directions can be treated as `constants' which commute with the Grassmann elements. Thus $\dot{q}_B$ is  a bosonic complex Grassmann matrix over the field of quaternionic numbers. These four directions will eventually be identified with emergent 4-D space-time;  with the real direction being time coordinate and the other three being spatial directions. We will often refer to these four directions as the `lower-half octonion plane', in contrast to the remaining four directions $(e_3, e_5, e_6, e_7)$ which we will refer to as the `upper half octonion plane'. The nomenclature is quite natural, from the viewpoint of our Lagrangian, as the lower plane has the character of a real line, and the upper plane the character of the imaginary line of a complex plane. If we look at the Top-right 4x4 plane, which comes from multiplying entries from the lower-half-plane and the upper-half-plane, it has all entries from only $(e_3, e_5, e_6, e_7)$. It is just like getting a pure imaginary number when we multiply a real number with a pure imaginary number. The same is true of the sub-table Bottom-left. We will assign the dynamical variable $q_B$, which describes Yang-Mills fields in our Lagrangian, the upper-half plane directions $(e_3, e_5, e_6, e_7)$. As if they were the imaginary part of the unified force $(\dot {q}_B, q_B)$. In fact, in our earlier paper \cite{MPSingh} the Yang-Mills fields were introduced precisely as the imaginary counterpart to gravity. The fact that this gels exactly with the octonion is encouraging. Together, $(\dot{q}_B, q_B)$ form an octonionic bosonic Grassmann matrix. The entries in the Bottom-right table are all from the lower-half plane $(e_0, e_1, e_2, e_4)$, which is reasonable: they have come from squaring the top-right: square of an imaginary number is real. This is also what puts the square $q_B^2$ in the `real / self-adjoint' part of the 8x8 matrix, as desired, even though $q_B$ itself is pure imaginary. Thus the bosonic part lies entirely in the Top-left. The Top-left and Bottom-right sub-tables are `real' and the Top-right and Bottom-left are `imaginary'. {\it We will now see that the fermions beautifully span both the real and imaginary directions.  This is what makes the Lagrangian non-self-adjoint, a property that turns out to be absolutely essential for unification! This is a convincing reason why the fundamental Lagrangian at the Planck scale must not be self-adjoint. In turn, the non-self-adjoint part of the Lagrangian / Hamiltonian, when significant, gives rise to emergence of the classical world via the Ghirardi-Rimini-Weber process of spontaneous localisation. Objective wave-function collapse models are an unavoidable consequence of the unification of interactions.}
 
 We will assume $q_F$ to be four-dimensional, and $\dot{q}_F$ to be also four-dimensional. Together they form an octonionic fermionic Grassmann matrix, and their respective directions are chosen, for the purpose of the present discussion,  in the following not-so-obvious manner, though the reason for this choice will become clear in the next sub-section. Similarly, $q_F^\dagger$ and $\dot{q}_F^\dagger$ form an octonionic Grassmann matrix with the shown directions. Thus we label the eight fermions as having the following eight linearly independent octonionic directions: $(V_\nu, V_{ad1}, V_{ad2}, V_{ad3}, V_{e+},  V_{u1},  V_{u2}, V_{u3})$ which stand respectively for the neutrino, the three anti-down quarks, the positron, and the three up quarks. The anti-particle directions are simply the complex conjugate of the corresponding particle direction [not the self-adjoint]. Thus,
 $(V^*_{a\nu}, V^*_{d1}, V^*_{d2}, V^*_{d3}, V^*_{e-}, V^*_{au1},  V^*_{au2}, V^*_{au3})$ denote respectively the anti-neutrino, three down quarks, the electron and three anti-up quarks.
 \begin{equation}
 \begin{split}
 \beta_1  \dot{q}_F^\dagger =  \bigg( V_\nu, V_{ad1}, V_{ad2}, V_{ad3} \bigg);   \qquad{\rm  [Neutrino,\  Anti-down\ quarks]} \\
    \beta_2 \dot{q}_F = \bigg(V_{e+}, V_{u1}, V_{u2}, V_{u3}\bigg) \qquad {\rm [Positron, \ Up \ quarks]}\\
  \beta_1  {q}_F^\dagger=   \bigg(V^*_{a\nu},V^*_{d1}, V^{*}_{d2}, V^{*}_{d3}\bigg); \qquad  {\rm [Anti-neutrino,\  Down\ quarks]}\\
 \beta_2 {q}_F = \bigg(V^*_{e-}, V^*_{au1}, V^{*}_{au2}, V^{*}_{au3}  \bigg) \qquad  {\rm [Electron, \ Anti-up \ quarks]}
 \end{split}
 \label{fermass}
\end{equation}
The names of the particles shown will be justified in the next sub-section.  We note the peculiarity that the anti-particles are obtained by taking the complex conjugate, and not by taking the adjoint of the particle. Thus the Higgs is perhaps composite of several [particle, octonionic-conjugate-of-anti-particle] pairs. It is not clear to us what implication the presence of the octonionic conjugate might have, in such constitution of the Higgs.
 
 \subsection{Using division algebras to relate the Lagrangian  to the standard model}
 \subsubsection {The overall picture} 
 
 The aikyon lives in an 8-D octonionic coordinate system, and the dynamics is described by complex-valued  Grassmann matrices. These matrices have matrix-valued components in the 8-D octonionic coordinate system. Thus, the bosonic matrix $\dot{q}_B$ introduced above and having components along the quaternion directions $(e_0, e_1, e_2, e_4)$ can be written as $\dot{q}_B = \dot{q}_{Be0} \; e_0 + \dot{q}_{Be1} \; e_1 + \dot{q}_{Be2}\ e_2 + \dot{q}_{Be4}\; e_4$. The components along the other four directions are zero. Similarly, the bosonic matrix $q_B$ has components along $(e_3, e_5, e_6, e_7)$ and can be written as $q_B = q_{Be3}\; e_3 + q_{Be5} \; e_5 + q_{Be6}\; e_6 + q_{Be7}\; e_7$. Together, they form the octonionic-coordinate based bosonic Grassmann matrix $( \dot{q}_{Be0} \; e_0 + \dot{q}_{Be1} \; e_1 + \dot{q}_{Be2}\ e_2 + \dot{q}_{Be4}\; e_4 + q_{Be3}\; e_3 + q_{Be5} \; e_5 + q_{Be6}\; e_6 + q_{Be7}\; e_7)$. The components of these eight matrices are even-grade complex Grassmann numbers. A similar interpretation holds for the fermionic odd-grade matrices present in the Lagrangian above.
 
 The automorphisms of the octonion algebra form the group $G_2$, which is the smallest of the exceptional Lie groups, and which has fourteen generators. From these generators, one can construct unitary transformations, and these unitaries, when they act on individual dynamical variables, leave the trace Lagrangian unchanged, as is also known in the theory of trace  dynamics [because of the allowed cyclic permutations inside the trace]. To possibly see this in another way, we recall from our earlier work  that the above Lagrangian can be brought to the following form [upto a total time derivative] after a redefinition of variables \cite{MPSingh}: 
\[
\mathcal{L} = Tr \biggl[\biggr. \dfrac{L_{p}^{2}}{L^{2}} \biggl(\biggr. \dot{\widetilde{Q}}_{B} + \dfrac{L_{p}^{2}}{L^{2}} \beta_{1} \dot{\widetilde{Q}}_{F}^\dagger\biggl.\biggr) \biggl(\biggr. \dot{\widetilde{Q}}_{B} + \dfrac{L_{p}^{2}}{L^{2}} \beta_{2} \dot{\widetilde{Q}}_{F} \biggl.\biggr) \biggl.\biggr]
\label{fulllagfund}
\]
where
\begin{equation}
{\dot{\widetilde{Q}}_B} = \frac{1}{L} (i\alpha q_B + L \dot{q}_B); \qquad  {\dot{\widetilde{Q}}_F} = \frac{1}{L} (i\alpha q_F + L \dot{q}_F);
\end{equation}
This trace Lagrangian can also be usefully written as
\[
\mathcal{L} = Tr \biggl[\biggr. \dfrac{L_{p}^{2}}{L^{2}} \biggl\{\biggr. \dot{\widetilde{Q}}_{B}^2   + \dfrac{L_{p}^{2}}{L^{2}} \dot{\widetilde{Q}}_{B}\biggl(\biggr.\beta_{1} \dot{\widetilde{Q}}^\dagger_{F} + +\beta_{2} \dot{\widetilde{Q}}_{F}  \biggl.\biggr)  + \dfrac{L_{p}^{4}}{L^{4}} \beta_{1} \dot{\widetilde{Q}}^\dagger_{F}   \beta_{2} \dot{\widetilde{Q}}_{F} \biggl.\biggr\} \biggl.\biggr]
\label{eq:tracelag}
\]
This makes the unification of interactions manifest. The first term is the bosonic part of the Lagrangian, the second term is the action of the bosons on the fermions, and the third term, the fermionic kinetic energy, is central for us to make contact with division algebras, and possibly also represents the Higgs boson. Spontaneous localisation, when it happens, separates gravitation, and its action on the fermions, from the internal symmetries. The weak symmetry separates from space-time and gravity, and combines with electromagnetism to form the electroweak symmetry. We call the entity described by the above Lagrangian, an `atom of space-time-matter' or an aikyon. The known elementary particles (quarks and leptons) and the gauge bosons, are special cases of the aikyon. Each of the three terms in this Lagrangian has the form of an octonion acting on itself, which helps understand why minimal left ideals of their algebra and the associated Clifford algebras are so important for understanding the standard model. 

The trace Lagrangian above can be written even more compactly as
\begin{equation}
\mathcal{L} =  Tr \biggl[\biggr. \dfrac{L_{p}^{2}}{L^{2}}  \dot{\widetilde{Q}}_{1sed}\;  \dot {\widetilde{Q}}_{2sed} \biggr]
\end{equation}
and where 
\begin{equation}
\dot{\widetilde{Q}}_{1sed}    =   \dot{\widetilde{Q}}_{B} + \dfrac{L_{p}^{2}}{L^{2}} \beta_{1} \dot{\widetilde{Q}}_{F}^\dagger  ; \ \qquad \dot {\widetilde{Q}}_{2sed} =  \dot{\widetilde{Q}}_{B} + \dfrac{L_{p}^{2}}{L^{2}} \beta_{2} \dot{\widetilde{Q}}_{F}
\end{equation}
Each of the two dynamical variables is now a sedenion. We see here the power of the aikyon concept: it unifies bosons and fermions and has a very simple Lagrangian which captures the standard model as well as gravity, and possibly also the four Higgs bosons. We will soon see that this Lagrangian actually already describes three fermion generations. For now however we will continue to work with the Lagrangian (\ref{fulllag2}) as it makes it easier to make contact with known physics.

 From the work of Cartan (as quoted in \cite{Ilka}, p. 923), it is known that the Lie algebra of $G_2$ possesses a symmetric invariant bilinear form
 \begin{equation}
 \beta \coloneqq x_0^2 + x_1 y_1 + x_2 y_2 + x_3 y_3
 \label{bina} 
 \end{equation}
 This is the form left invariant by the lowest dimension  representation of this Lie algebra, which happens to be a seven-dimensional complex representation. The scalar product $\beta$ has real coefficients, which is consistent with the presence of the real coefficients $\alpha$ and $L$ in our trace Lagrangian. 
 Although we do not have a proof for this yet, the invariance of the trace Lagrangian under unitaries made from the generators of $G_2$ suggests that this trace is also Cartan's invariant bilinear form. As in known in trace dynamics, the existence of the Adler-Millard conserved charge is a consequence of this global unitary invariance, and this charge perhaps also has an intimate connection with the Lie algebra of $G_2$.

 Unitary transformations of the dynamical variables are analogs of general coordinate transformations (in the Riemannian geometry of general relativity) in the present context of the 
 non-commuting octonionic coordinates. The invariance of the trace Lagrangian, and the observation that this trace Lagrangian describes [quantum] gravity and the standard model (to be shown below) motivates us to make the following proposal. Physical laws are invariant under automorphisms [`general coordinate transformations'] of the octonionic coordinates, and this is responsible for the emergence of interactions as the (non-commutative)  geometry of the octonionic space. We  can elaborate on the apparent similarity of this assertion to  the laws of special and general relativity, via the following observations. The extra four dimensions that are added on to space-time arise very naturally in our trace dynamics Lagrangian \cite{Singhspin}, and are analogous to the extra dimensions in Kaluza-Klein theories and give rise to the standard model forces. With the following difference from Kaluza-Klein theories: the entire 8-D space is now a non-commutative octonionic space, and the theory does not have to be quantised.  Rather, because it is a trace dynamics, quantum (field) theory is emergent from our theory. The extra dimensions get suppressed in the classical limit [as we show below] because the fermions undergo spontaneous localisation under suitable conditions, and get localised to the real 4-D part of the octonionic space. Thus the symmetry group of the unified theory is $G_2$, which unifies the Lorentz group  with the standard model symmetry group $SU(3)_C \times SU(2)_L \times U(1)_Y$. We will explain how gravity emerges only in the classical limit, as a consequence of spontaneous localisation of a large collection of aikyons. The octonionic space is the non-commutative analog of a space-time manifold. Every quantum elementary particle has its own such space - this describes the small-scale structure of space-time. But one does not make a distinction between the particle and the octonionic space-time which it inhabits. Different aikyons interact via entanglement, a feature more fundamental than quantum theory. The latter inherits the property of entanglement from the underlying trace dynamics.

 Automorphisms mix bosonic and fermionic terms in the trace Lagrangian. We recall though \cite{Singhspin}  that in our trace dynamics, bosonic matrices and fermionic matrices do not have a fixed spin (there is no Planck's constant in the underlying theory, it being only emergent, and the integral / half-integral spin of bosons / fermions is also emergent). So the mixing of the two kinds of terms is not a problem. However, assuming that the bilinear form (\ref{bina}) exists, it is always possible to write the Lagrangian in our chosen form, and then the following concrete interpretation of the various terms exists, leading to the association of this Lagrangian with the standard model.
 
 We now work out the directions and components for each of the eight terms inside  the trace Lagrangian. The first term  $T_1$ proportional to $\dot{q}_B^2$ comes from squaring of 
 $\dot{q}_B = \dot{q}_{Be0} \; e_0 + \dot{q}_{Be1} \; e_1 + \dot{q}_{Be2}\ e_2 + \dot{q}_{Be4}\; e_4$:
 \begin{equation} 
 \begin{split}
 \dot{q}_B^2  & = \bigg [  \dot{q}_{Be0} \; e_0 + \dot{q}_{Be1} \; e_1 + \dot{q}_{Be2}\ e_2 + \dot{q}_{Be4}\; e_4 \bigg ] \times \bigg [  \dot{q}_{Be0} \; e_0 + \dot{q}_{Be1} \; e_1 + \dot{q}_{Be2}\ e_2 + \dot{q}_{Be4}\; e_4 \bigg ] \\  & =  \dot{q}_{Be0}^2 - \dot{q}_{Be1}^2 -\dot{q}_{Be2}^2 -\dot{q}_{Be4}^2 + 2 \dot{q}_{Be0} \; \big[ \dot{q}_{Be1} \; e_1 + \dot{q}_{Be2}\ e_2 + \dot{q}_{Be4}\; e_4 \big]
 \label{bofi}
 \end{split}
 \end{equation}
 The other cross-terms in the bi-product mutually cancel because of the product rule for octonions, and because in the trace Lagrangian, product of two bosonic Grassmann matrices commutes. The various terms in the resulting last line have the following interpretation, as we will show in detail below. It will be shown that $\dot{q}_B^2$ describes Lorentz symmetry 
 {\it and} the weak symmetry of the standard model! Indeed, these two symmetries arise together from the underlying theory, and are separated only when spontaneous localisation leads to the emergence of classical space-time. One might think of spontaneous localisation as a kind of spontaneous symmetry breaking which results from large-scale entanglement of fermions. The weak symmetry then becomes one of the three internal symmetries of the standard model. Thus the first four terms in the last line above describe  the Lorentz symmetry and gravitation of a 4-D space-time: these terms will become the Einstein-Hilbert action in the classical limit, because they constitute $Tr [D_B^2]$, where $D_B$ is the conventional Dirac operator. But the negative squared terms also describe the weak isospin bosons, highlighting the connection between gravity and the weak force, which we discuss in some detail below. The last term describes the space-time coupling of the three weak bosons, and is possibly connected with the Higgs mechanism which will give rise to mass to the  weak isospin fields.
 
 The presence of 4-D Lorentz symmetry, despite there being non-commuting quaternionic directions present, could help us understand how quantum non-locality operates. The `space-time' directions do not commute with each other, so there cannot be locality in the sense of disallowing influence outside the light cone. Nonetheless, because Lorentz symmetry is present, relativistic quantum field theory is consistent with special relativity. In the classical limit, space-time directions commute with each other, to an excellent approximation, and causal light-cone structure emerges. 
 
 We note that the matrix trace is to be taken as usual, keeping the octonion direction fixed. This will yield `imaginary octonion direction based' terms in the trace Lagrangian as well as in the action, apart from the real terms. These imaginary terms go away in the emergent quantum theory and in the classical limit, leaving behind the desired Lagrangian. However, the imaginary terms are absolutely essential for describing unification in the underlying trace dynamics. 
 
 We now work out the components of the second term $T_2$ in the Lagrangian, which is proportional to $\dot{q}_B \beta_2 \dot{q}_F$, using the coordinate assignment for the fermions shown in Eqn. (\ref{fermass}). We have
 \begin{equation}
 \begin{split}
  \dot{q}_B  \beta_2 \dot{q}_F   =  \big [  \dot{q}_{Be0} \; e_0 + \dot{q}_{Be1} \; e_1 + \dot{q}_{Be2}\ e_2 + \dot{q}_{Be4}\; e_4 \big ] \times\\
   \bigg [ \dot{q}_{Fe+} \;  V_{e+}   + \dot{q}_{Fu1} \; \frac{1}{4}  V_{u1} + \dot{q}_{Fu2} \; V_{u2}  + \dot{q}_{Fu3} \; V_{u3} \bigg]\\
 = \big [  \dot{q}_{Be0} \; e_0  ] \times  \bigg [ \dot{q}_{Fe+} \;  V_{e+}   + \dot{q}_{Fu1} \; V_{u1}  + \dot{q}_{Fu2} \; V_{u2}  + \dot{q}_{Fu3} \; V_{u3} \bigg] +\\
 \big [   \dot{q}_{Be1} \; e_1  \big ] \times  \bigg [ \dot{q}_{Fe+} \;  V_{e+}   + \dot{q}_{Fu1} \; V_{u1}  + \dot{q}_{Fu2} \; V_{u2}  + \dot{q}_{Fu3} \; V_{u3}\bigg] +\\
\big [  \dot{q}_{Be2}\ e_2  \big ] \times \bigg [ \dot{q}_{Fe+} \;    V_{e+} + \dot{q}_{Fu1} \; V_{u1}  + \dot{q}_{Fu2} \; V_{u2}  + \dot{q}_{Fu3} \; V_{u3} \bigg] +\\
  \big [   \dot{q}_{Be4}\; e_4 \big ] \times \bigg [ \dot{q}_{Fe+} \;   V_{e+}  + \dot{q}_{Fu1} \; V_{u1}  + \dot{q}_{Fu2} \; V_{u2}  + \dot{q}_{Fu3} \; V_{u3} \bigg]\\
 \end{split}
 \end{equation}
 These 16 terms have a simple interpretation. The first four represent what is usually referred to as Dirac-spin; the action of the conventional Dirac operator on four  [positron, up quarks] of the eight fermions of a generation. (The other four fermions will come from the next term $T_3$ in the Lagrangian). The remaining 12 terms above describe the coupling of the three weak bosons to the four fermions. Remarkably,  after spontaneous localisation leads to the emergence of classical space-time,  the Dirac-spin terms separate from the rest, and give rise to the classical action for the relativistic point particle in general relativity. As a result,  the weak interaction  separates, and emerges as an internal symmetry, precisely because the directions $(e_1, e_2, e_4)$ along which the weak bosons lie are  imaginary octonionic directions.  
 
 In a way similar to the term $T_2$, we can now expand the third term $T_3$ of the Lagrangian, which is proportional to $\dot{q}_B \beta _1 \dot{q}^\dagger_F$. We have brought $\dot{q}_B$ to the front, which can be done inside the trace, and without a change of sign, because $\dot{q}_B$ is bosonic. Thus we have, 
 \begin{equation}
 \begin{split}
  \dot{q}_B \beta _1 \dot{q}^\dagger_F= \big [  \dot{q}_{Be0} \; e_0 + \dot{q}_{Be1} \; e_1 + \dot{q}_{Be2}\ e_2 + \dot{q}_{Be4}\; e_4 \big ] \times \\  \bigg[  q_{F\nu}\;V_\nu + q_{Fad1\;} V_{ad1} + q_{Fad2} \;V_{ad2} + q_{Fad3}\; V_{ad3} \bigg]=\\
   \big [  \dot{q}_{Be0} \; e_0 \big ] \times \bigg[  q_{F\nu}\;V_\nu + q_{Fad1\;} V_{ad1} + q_{Fad2} \;V_{ad2} + q_{Fad3}\; V_{ad3} \bigg] +\\
   \big [  \dot{q}_{Be1} \; e_1  \big ] \times \bigg[  q_{F\nu}\;V_\nu + q_{Fad1\;} V_{ad1} + q_{Fad2} \;V_{ad2} + q_{Fad3}\; V_{ad3} \bigg] + \\
   \big [   \dot{q}_{Be2}\ e_2 \big ] \times \bigg[  q_{F\nu}\;V_\nu + q_{Fad1\;} V_{ad1} + q_{Fad2} \;V_{ad2} + q_{Fad3}\; V_{ad3} \bigg]+ \\ 
    \big [  \dot{q}_{Be4}\; e_4 \big]\times  \bigg[  q_{F\nu}\;V_\nu + q_{Fad1\;} V_{ad1} + q_{Fad2} \;V_{ad2} + q_{Fad3}\; V_{ad3} \bigg]
 \end{split}
 \end{equation}
 In analogy with the term $T_2$, this term describes the action of the Dirac operator on the other four fermions of a generation [neutrino, anti-down quarks], and the action of the three weak bosons on these four fermions. Thus, together the terms $T_1, T_2, T_3$  of the Lagrangian describe the unified Lorentz-Weak symmetry. We will see below that this symmetry is described by the Clifford algebra $C\ell (6)$. This result is already there in the beautiful work of Stoica \cite{Stoica}, and we will essentially repeat and report his, and Furey's,  work below, in the context of our theory.
 
 The other three terms of the Lagrangian, $T_5, T_6, T_7$ which we now analyse, analogously are associated with another copy of the Clifford algebra $C\ell(6)$. These describe the strong and electromagnetic interactions of quarks and leptons, via $SU(3)_c \times U(1)_{em}$, as shown by Furey \cite{f1}. Why, one might ask, does this copy of $C\ell(6)$ behave differently from the first one? The answer lies in the fact that an octonion has only one real direction, not two. The first set of bosons associate with the directions $(e_0, e_1, e_2, e_4)$ of which $e_0$ is real. The second set of bosons (to be analysed below) - the gluons and the photon - associate with the directions $(e_3, e_5, e_6, e_7)$, all of which are imaginary. The presence of the real direction $e_0$ in the first set is what causes spontaneous localisation to occur, which results in the emergence of classical space-time and gravitation, and its separation from the three emergent internal symmetries.
 
 Let us now look at term $T_5$, which is proportional to $q_B^2$. We recall that the assigned octonion directions for $q_B$ are $(e_3, e_5, e_6, e_7)$. Hence we get that
 \begin{equation} 
 \begin{split}
 {q}_B^2  & = \bigg [  {q}_{Be3} \; e_3 + {q}_{Be5} \; e_5 + {q}_{Be6}\ e_6 + {q}_{Be7}\; e_7 \bigg ] \times \bigg [  {q}_{Be3} \; e_3 + {q}_{Be5} \; e_5 + {q}_{Be6}\ e_6 + {q}_{Be7}\; e_7\bigg ] \\  & =  -{q}_{Be3}^2 - {q}_{Be5}^2 -{q}_{Be6}^2 -{q}_{Be7}^2  \end{split}
 \end{equation}
 All the cross terms cancel, because of the multiplication rule for the octonions. The four resulting terms, which describe the photon and gluons of three colours, should be compared and contrasted with the first set of bosonic terms derived in Eqn. (\ref{bofi}). The difference is obvious: the first term there is positive, and there is also a cross-term, because of the presence of the real direction $e_0$.
 
 Next, let us look at the cross-terms $T_6 = q_B \beta_2 q_F  $ and $T_7 = q_B \beta_1 q_F^\dagger$. These describe the action of gluons and the photon on the quarks and leptons:
 \begin{equation}
 \begin{split}
  {q}_B  \beta_2 {}{q}_F   =  \big [  {}{q}_{Be3} \; e_3 + {}{q}_{Be5} \; e_5 + {}{q}_{Be6}\ e_6 + {}{q}_{Be7}\; e_7 \big ] \times\\
   \bigg [ {q}_{Fe-} \;  V^*_{e-}   + {}{q}_{Fau1} \;  V^{*}_{au1}  + {}{q}_{Fu2} \; V^{*}_{au2}  + {}{q}_{Fu3} \; V^{*}_{au3} \bigg]\\
 = \big [  {}{q}_{Be3} \; e_3  ] \times  \bigg [ {}{q}_{Fe-} \;  V^*_{e-}   + {}{q}_{Fau1} \; V^{*}_{au1}  + {}{q}_{Fau2} \; V^*_{au2}  + {}{q}_{Fau3} \; V^{*}_{au3} \bigg] +\\
 \big [   {}{q}_{Be5} \; e_5  \big ] \times  \bigg [ {}{q}_{Fe-} \;  V^*_{e-}   + {}{q}_{Fau1} \; V^{*}_{au1}  + {}{q}_{Fau2} \; V^{*}_{au2}  + {}{q}_{Fau3} \; V^{*}_{au3}\bigg] +\\
\big [  {}{q}_{Be6}\ e_6  \big ] \times \bigg [ {}{q}_{Fe-} \;  V^{*}_{e-}   + {}{q}_{Fau1} \; V^{*}_{au1}  + {}{q}_{Fau2} \; V^{*}_{au2}  + {}{q}_{Fau3} \; V^{*}_{au3}+ \bigg] +\\
  \big [   {}{q}_{Be7}\; e_7\big ] \times \bigg [ {}{q}_{Fe-} \;  V^{*}_{e-}   + {}{q}_{Fau1} \; V^{*}_{au1}  + {}{q}_{Fau2} \; V^{*}_{au2}  + {}{q}_{Fau3} \; V^{*}_{au3} \bigg]\\
 \end{split}
 \end{equation}
Out of the four components $  ( {q}_{Be3} \; e_3 + {q}_{Be5} \; e_5 + {q}_{Be6}\ e_6 + {q}_{Be7}\; e_7 ) $, the one representing the photon will be moved, along with terms representing its action on the fermions,  to join the weak interaction terms, after spontaneous localisation separates the Dirac operator terms from the weak interaction terms. Thus, $U(1)_{em}$ joins with $SU(2)_W$ to form $SU(2)_L \times U(1)_Y$.

  \begin{equation}
 \begin{split}
  {}{q}_B \beta _1 {}{q}^\dagger_F= \big [  {}{q}_{Be3} \; e_3 + {}{q}_{Be5} \; e_5 + {}{q}_{Be6}\ e_6 + {}{q}_{Be7}\; e_7 \big ] \times \\  \bigg[ q_{Fa\nu}\;V^*_{a\nu} + q_{Fd1\;} V^*_{d1} + q_{Fd2} \; V^*_{d2} + q_{Fd3}\; V^*_{d3}  \bigg]=\\
   \big [  {}{q}_{Be3} \; e_3 \big ] \times \bigg[  q_{Fa\nu}\;V^*_{a\nu} + q_{Fd1\;} V^*_{d1} + q_{Fd2} \; V^*_{d2} + q_{Fd3}\; V^*_{d3} \bigg] +\\
   \big [  {}{q}_{Be5} \; e_5  \big ] \times \bigg[  q_{Fa\nu}\;V^*_{a\nu} + q_{Fd1\;} V^*_{d1} + q_{Fd2} \;V^*_{d2} + q_{Fd3}\; V^*_{d3} \bigg] + \\
   \big [   {}{q}_{Be6}\ e_6 \big ] \times \bigg[  q_{Fa\nu}\;V^*_{a\nu} + q_{Fd1\;} V^*_{d1} + q_{Fd2} \; V^*_{d2} + q_{Fd3}\; V^*_{d3} \bigg]+ \\ 
    \big [  {}{q}_{Be7}\; e_7 \big]\times  \bigg[ q_{Fa\nu}\;V^*_{a\nu} + q_{Fd1\;} V^*_{d1} + q_{Fd2} \; V^*_{d2} + q_{Fd3}\; V^*_{d3} \bigg]
    \label{cross}
 \end{split}
 \end{equation}

Lastly, we look at the terms $T_4$ and $T_8$ in the Lagrangian, which can be together written as
\begin{equation}
\begin{split}
& T_4 + T_8 = \frac{L_P^4}{L^4}\bigg [  \beta_1\dot{q}^\dagger_F \beta_2\dot{q}_F  -\frac{\alpha^2}{L^2} \beta_1 q^\dagger_F \beta_2 q_F  \bigg] =\\ & \frac{1}{8}\frac{L_P^4} {L^4} V_{e+} \bigg [\bigg( \dot{q}_{\nu} \dot{q}_{Fe+} + \dot{q}_{Fad1}\dot{q}_{Fu1}  + \dot{q}_{Fad2}\dot{q}_{Fu2}  + \dot{q}_{Fad3}\dot{q}_{Fu3}\bigg)  - \\ & \frac{\alpha^2}{L^2}\bigg(q_{Fa\nu}q_{Fe-}  + q_{Fd1} q_{Fau1}+ q_{Fd2} q_{Fau2} +  q_{Fd3} q_{Fau3}\bigg)  \bigg]
\end{split}
\label{higgs}
\end{equation}
Here we see the action of fermions onto themselves. The bi-octonion structure is missing here, because we are not retaining total time derivatives in the Lagrangian. However the bi-octonionic structure is evident in the form (\ref{fulllagfund}) of the Lagrangian.
It remains to be understood if these terms can provide one of the four required Higgs bosons. They do possess the form of a kinetic energy term minus a potential energy term.

Having introduced the various terms in the Lagrangian, we now justify their claimed relation with the standard model. For the most part, the work on division algebras and Clifford algebras by Furey \cite{f1}, Stoica \cite{Stoica}  and several other researchers, essentially accomplishes this already. We report their work below, to justify the connection of our Lagrangian with division algebras and the standard model. The new part is the direct evidence for the presence of the gravito-weak symmetry in our Lagrangian, whose existence has been conjectured by several researchers, notably by Onofrio \cite{Onofrio1, Onofrio2}, \cite{Percacci, Percacci2},  and also hinted at by Stoica \cite{Stoica}.

\bigskip

\subsubsection  {The Lorentz symmetry and the Clifford algebra $C\ell (2)$}
If we consider the four quaternion components of $\dot{q}_B$, whose bi-product lies in the Top-Left part of the multiplication table in Fig. 1, we recall that the term $T_1 \propto \dot{q}_B^2$ is given by Eqn. (\ref{bofi}). For the special case that each of the four bosonic components is equal to the identity matrix, the sum is of the form $I \times (1, -1, -1, -1)$. This form is Lorentz invariant and the Lorentz algebra $SL(2,C)$ is known to be generated by complex quaternions. Because it aids understanding the Clifford algebra $C\ell(6)$ we recall that complex quaternions give a faithful representation of $C\ell(2)$. This can be arrived at using the following two fermionic ladder operators:
\begin{equation}
\alpha_0 = \frac{1}{2} (ie_1 - e_2) ; \qquad \alpha_0^\dagger =\frac{1}{2} (ie_1 + e_2);  \qquad \alpha_0^2 = \alpha_0^{\dagger 2} = 0,  \{\alpha_0, \alpha_0^\dagger \} = 1
\label{a0}
\end{equation}
Spinors can be constructed as minimal left ideals of $C\ell(2)$ by using the idempotent 
$V\equiv\alpha \alpha^\dagger$ which acts like the vacuum state. Left multiplying $V$ by the Clifford algebra of complex quaternions is the minimal left ideal, which happens to be a 2-D complex space spanned by $V$ and $\alpha^\dagger V$. Their linear combination gives left-handed Weyl spinors under the Lorentz algebra $SL(2,C)$. Similarly, $C\ell(2)$ acting on the complex-conjugate $V^{*}$ gives rise to the space $(V^{*}, \alpha V^{*})$ whose linear combination gives rise to right-handed  Weyl spinors \cite{f1}. The left-handed Weyl spinors and the right-handed Weyl spinors are simply complex conjugates of each other. 

This construction, though elementary, has deep significance and implications for our theory, considering that Furey \cite{f1} then develops a completely analogous construction for the Clifford algebra $C\ell(6)$, from the octonion algebra, to describe quarks and leptons and their unbroken $SU(3)_c \times U(1)_{em}$ symmetry. This puts the Weyl spinors on the same footing as the fermions, as if the former were particles too. And that is how it turns out to be, as we will soon see. This also reinforces the aikyon concept, which puts space-time and matter on the same footing, at the Planck scale. This interpretation is further strengthened because Stoica \cite{Stoica} incorporates the Lorentz algebra and the weak symmetry together, to form the other copy of $C\ell(6)$, when describing the symmetries of quarks and leptons in the standard model. This leads us to propose the gravito-weak interaction.

\bigskip

\subsubsection{The unbroken electro-colour symmetry and the Clifford algebra $C\ell(6)$}
This section follows \cite{f1}, and is a (condensed) repetition of the results therein, to put those results in the context of our theory, and to link up to the next sub-section. The reader is referred to \cite{f1} for further details on arriving at electro-colour symmetry, from the algebra of octonions.

The standard model symmetries are being arrived at by multiplying the algebra onto itself. But why does this scheme work? The answer becomes obvious when we take note that in our Lagrangian, every term is quadratic or bilinear. Therefore when we compute the terms of the Lagrangian, written out in octonionic space, the algebra inevitably acts on itself, and that dictates the symmetries. The Grassmann matrices simply go for a ride when the symmetries of the Lagrangian are being determined, and the role of the matrices comes to the fore in the dynamics and in the emergent theory. The algebra has no further role to play in the emergent quantum theory, nor in the classical limit. But by then the algebra has already made its mark, and the correctly predicted properties of the standard model are evidence that division algebras are at work at the Planck scale.

Multiplication of octonions is not associative, whereas Clifford algebras are associative. Hence the former cannot give a faithful representation of the latter. Nonetheless, there is a trick using which $C\ell(6)$ can be built from the complex octonions. The idea is to think of a chain of left multiplications in the (non-associative) algebra as a way to generate maps from one element in the algebra to another. And maps are necessarily associative. For the complex quaternions, this yields nothing new: an associative algebra leads to maps which are necessarily associative. However, for complex octonions this works wonders, thanks to their amazing mathematical properties. It turns out that to describe the most general chain of left multiplying elements of the algebra [and these chains then serve as maps which are associative], one does not need chains of length more than three, and these chains, acting as maps, behave as elements of a Clifford algebra. It is easily shown that there are sixty-four such independent complex-valued octonionic chains [one of length zero, seven of length one, twenty-one of length two, and thirty-five of length three]. They can be thought of as being equivalent to 8x8 complex matrices, and give a faithful representation of $Cl(6)$. 

Then, just like in the case of $C\ell(2)$ above, where the complex quaternions were used to construct a new basis, a new basis is constructed now, from one-vectors, noting that unity is the zero vector, and product of any six imaginary directions, say $(e_1, e_2, e_3, e_4, e_5, e_6)$ gives the seventh one, in this case $e_7$. So the six imaginary directions are used to make six fermionic  ladder operators, and these are the following:
\begin{equation}
\begin{split}
\alpha_1 = \frac{1}{2} [-e_5 + ie_4]; \qquad \alpha_2 = \frac{1}{2} [-e_3 + ie_1]; \qquad \alpha_3 = \frac{1}{2} [-e_6 + ie_2]; \\
\alpha_1^\dagger =  \frac{1}{2} [e_5 + ie_4]; \qquad \alpha_2^\dagger =  \frac{1}{2} [e_3 + ie_1]; \qquad \alpha_3^\dagger =  \frac{1}{2} [e_6 + ie_2]
\label{ladelec}
\end{split}
\end{equation}
it being understood that all multiplications are left multiplications. These ladder operators satisfy the Clifford algebra
\begin{equation}
\{\alpha_i, \alpha_j \} = 0; \qquad \{\alpha_i^\dagger, \alpha_j^\dagger\} =0; \qquad \{\alpha_i, \alpha_j^\dagger\} = \delta_{ij}
\label{ladec}
\end{equation}
The number operator is defined as usual
\begin{equation}
N = \sum_1^3 \alpha_i^\dagger \; \alpha_i
\end{equation}
The ladder operators have a unitary symmetry $U(3)$, which rotates the lowering operators amongst themselves, and the raising operator amongst themselves. And we know that 
$U(3) = SU(3)\times U(1)/ \mathbb{Z}_3$. It so happens that $SU(3)$ is a sub-group of $G_2$ which holds one of the imaginary units constant. Whereas $U(1)$ is generated by the number operator $N$ above.

As before, one now constructs minimum left ideals of the Clifford algebra $C\ell(6)$, by first defining the idempotent $V$ (the projector), which acts as the `vacuum', as follows:
\begin{equation} 
V = \alpha_1 \alpha_2 \alpha_3 \alpha_3^\dagger \alpha_2^\dagger\alpha_1^\dagger = \frac{i}{2} e_7 \qquad [\rm Neutrino]
\end{equation}
The particular form $ie_7/2$ is of course specific to the coordinates we have chosen, and we shortly explain why this is the neutrino. The minimal left ideal is obtained by left multiplying on the vacuum $V$ by the Clifford algebra; this yields an eight complex dimensional space denoted $S^u$ and spanned by the following basis vectors:
\begin{equation}
\begin{split}
V = \frac{i}{2} e_7 \qquad [V_\nu  \  {\rm Neutrino}]\\
\alpha_1^\dagger V = \frac{1}{2} ( e_5 + ie_4)\times V = \frac{1}{4} ( e_5 + ie_4) \qquad [\rm V_{ad1}\  Anti-down\ quark] \\
\alpha_2 ^\dagger V = \frac{1}{2} ( e_3 + ie_1)\times V = \frac{1}{4} ( e_3 + ie_1) \qquad [\rm V_{ad2}\ Anti-down\ quark] \\
\alpha_3^\dagger V = \frac{1}{2} ( e_6 + ie_2)\times V = \frac{1}{4} ( e_6 + ie_2) \qquad [\rm V_{ad3} \ Anti-down\ quark] \\
\alpha_3^\dagger \alpha_2^\dagger V = \frac{1}{4} ( e_4+ ie_5) \qquad [\rm V_{u1}\ Up\ quark] \\
\alpha_1^\dagger \alpha_3^\dagger V =  = \frac{1}{4} ( e_1 + ie_3) \qquad [\rm V_{u2}\ Up\ quark] \\
\alpha_2^\dagger \alpha_1^\dagger V =  \frac{1}{4} ( e_2 + ie_6) \qquad [\rm V_{u3}\ Up\ quark] \\
\alpha_3^\dagger \alpha_2^\dagger \alpha_1^\dagger V = -\frac{1}{8}(i+e_7) \qquad [{\rm V_{e+}\ Positron}]
\end{split}
\end{equation}
Similarly, one can define the space $S_d$ by acting the algebra on the conjugate idempotent $V^*$, spanned by the following eight vectors:
\begin{equation}
\begin{split}
V^* = -\frac{i}{2} e_7 \qquad  [ V_{a\nu}\ \rm Anti-neutrino]\\
\alpha_1 V^* = \frac{1}{2} ( -e_5 + ie_4)\times V^* = \frac{1}{4} ( -e_5 + ie_4) \qquad [ V^*_{d1} \  {\rm Down\ quark}] \\
\alpha_2  V^* = \frac{1}{2} ( -e_3 + ie_1)\times V^* = \frac{1}{4} ( -e_3 + ie_1) \qquad [V^*_{d2}\ \rm Down\ quark] \\
\alpha_3 V^* = \frac{1}{2} ( -e_6 + ie_2)\times V^* = \frac{1}{4} ( -e_6 + ie_2) \qquad [V^*_{d3}\ \rm Down\ quark] \\
\alpha_2 \alpha_3 V^* = \frac{1}{4} ( -e_4+ ie_5) \qquad [V^*_{au1}\ \rm Anti-up\ quark] \\
\alpha_3 \alpha_1 V^* =  = \frac{1}{4} ( -e_1 + ie_3) \qquad [V^*_{au2}\ \rm Anti-up\ quark] \\
\alpha_1 \alpha_2 V^* =  \frac{1}{4} ( -e_2 + ie_6) \qquad [V^*_{au3}\ \rm Anti-up\ quark] \\
\alpha_1 \alpha_2 \alpha_3 V^* = \frac{1}{8}(i+e_7) \qquad [V^*_{e-}\ {\rm Electron}]
\end{split}
\end{equation}
The action of the algebra shows that anti-particles are simply complex conjugates of the particles. Now one needs to justify the particle labels in the above basis. The ladder operators transform under $U(3)$ symmetry [made of $SU(3)$ and $U(1)$)], and hence so do the vectors in the above basis. Consider the action of $SU(3)$ first. Now we have noted that $SU(3)$ is a sub-group of $G_2$ obtained by keeping one of the imaginary directions, say $e_7$, fixed. The fourteen generators of  $G_2$ can be expressed as chains of octonion elements acting on the octonions. One could ask how out of all of $G_2$, the groups $SU(3)$ and $U(1)$ are selected. It has been shown that the generating space of a Clifford algebra $C\ell(n)$ with $n$ even can always be partitioned into two maximal totally isotropic subspaces (MTIS) \cite{f1} and in the case of $C\ell(6)$ these are: one MTIS spans $(\alpha_1, \alpha_2, \alpha_3)$ and the other spans $(\alpha_1^\dagger, \alpha_2^\dagger, \alpha_3^\dagger)$. If this separation is to be preserved under transformations of the ladder operators, the MTIS are precisely those generated by the Lie algebra of $SU(3)$ and $U(1)$. The $SU(3)$ generators made from these ladder operators are given in Eqn. (6.26) of \cite{f1} and we do not repeat them here. The generator for $U(1)$ is given by $Q=N/3$ where $N$ is the number operator given above.

Now, under $SU(3)$, the state $V$ in $S^u$ translates as a singlet, the next three out of those translate as anti-triplets, the three after those as triplets, and the last one as a singlet. Analogously, in $S^d$, the first state $V^*$ transforms as a singlet, the next three as a triplet, the three after those as anti-triplets, and the last one as a singlet. Next, one finds the eigenvalues of the operator $Q$ and finds them to be $(0, 1/3, 1/3, 1/3, 2/3, 2/3, 2/3, 1)$ and the eigenvalues of the eight states listed under $S^u$ are precisely these values, in this very order. Analogously, in $S^d$, the charges for the eight states come out to be $(0, -1/3, -1/3, -1/3, -2/3, -2/3, -2/3, -1)$. We can now identify these objects given these properties. In $S^u$, a singlet under $SU(3)$ with zero charge is either a neutrino or an anti-neutrino: it turns out to be the neutrino as we justify in the next sub-section. The three anti-triplets with charge $1/3$ are anti-down quarks, the three triplets with charge $2/3$ are quarks, and the singlet with charge $1$ is the positron. Similarly, in $S^d$, one identifies an anti-neutrino, three down quarks, three anti-up quarks, and the electron.

One also needs to identify the eight gluons and the boson for $U(1)_{em}$. There being eight gluons with their corresponding generators, the four $q_B$ are not enough. This compels us to drop the assumption that the $q_B$ are self-adjoint. So we reconstruct the Lagrangian as
\begin{equation}
\begin{split}
 \frac{S}{C_0} = \frac{1}{2}\int \frac{d\tau}{\tau_{Pl}}\; Tr \bigg[\biggr.\frac{L_P^2}{L^2}\bigg\{\biggr.  \bigg(\dot{q}_B^2 +\frac{L_P^2}{L^2} \dot{q}_B\beta_2 \dot{q}_F + \frac{L_P^2}{L^2} \beta_1 \dot{q}^\dagger_F \dot{q}_B   + \frac{L_P^4}{L^4} \beta_1\dot{q}^\dagger_F \beta_2\dot{q}_F \bigg) \\- \frac{\alpha^2 }{L^2} \bigg(q_B^\dagger q_B     
+ \frac{ L_P^2}{L^2} q_B \beta_2 q_F + \frac{ L_P^2}{L^2} \beta_1 q^\dagger_F q_B^\dagger +  \frac{L_P^4}{L^4} \beta_1 q^\dagger_F  \beta_2 q_F  \bigg)
 \biggl. \bigg\} \biggl.\bigg] 
 \end{split}
 \label{fulllag2rep1}
 \end{equation} 
while still keeping $\dot{q}_B$ self-adjoint. Therefore we now have a total of eight $(q_B, q_B^\dagger)$ and they can be assigned the eight bosonic generators constructed by Furey for $SU(3)_c$ \cite{f1}. The cross terms (\ref{cross}) are replaced by
 \begin{equation}
 \begin{split}
  {}{q}_B^\dagger \beta _1 {}{q}^\dagger_F= \big [  {}{q}_{Be3} \; e_3 + {}{q}_{Be5} \; e_5 + {}{q}_{Be6}\ e_6 + {}{q}_{Be7}\; e_7 \big ] \times \\  \bigg[ q_{Fa\nu}\;V^*_{a\nu} + q_{Fd1\;} V^*_{d1} + q_{Fd2} \; V^*_{d2} + q_{Fd3}\; V^*_{d3}\bigg]=\\
   \big [  {}{q}_{Be3}^\dagger \; e_3 \big ] \times \bigg[  q_{Fa\nu}\;V^*_{a\nu} + q_{Fd1\;} V^*_{d1} + q_{Fd2} \; V^*_{d2} + q_{Fd3}\; V^*_{d3} \bigg] +\\
   \big [  {}{q}_{Be5}^\dagger \; e_5  \big ] \times \bigg[  q_{Fa\nu}\;V^*_{a\nu} + q_{Fd1\;} V^*_{d1} + q_{Fd2} \;V^*_{d2} + q_{Fd3}\; V^*_{d3} \bigg] + \\
   \big [   {}{q}_{Be6}^\dagger e_6 \big ] \times \bigg[  q_{Fa\nu}\;V^*_{a\nu} + q_{Fd1\;} V^*_{d1} + q_{Fd2} \; V^*_{d2} + q_{Fd3}\; V^*_{d3} \bigg]+ \\ 
    \big [  {}{q}_{Be7}^\dagger\; e_7 \big]\times  \bigg[ q_{Fa\nu}\;V^*_{a\nu} + q_{Fd1\;} V^*_{d1} + q_{Fd2} \; V^*_{d2} + q_{Fd3}\; V^*_{d3} \bigg]
 \end{split}
 \end{equation}
 The $U(1)_{em}$ boson stays as the number operator, as before.

It is a great triumph of Furey's work that she is able to derive the quantisation of electric charge, and three colours, for quarks and leptons, simply from the algebra of complex octonions acting on themselves. It is a strong hint that fundamentally the fermions and exchange bosons of the standard model  live in an octonionic space, and their dynamics is described by a Lagrangian with a quadratic form. The existence of such a dynamics is borne out by the Lagrangian we have constructed. Our Lagrangian was not constructed to explain the significance of division algebras. Rather, our goal stemmed from quantum foundations: to achieve a reformulation of quantum field theory which does not depend on classical time. This led us to the theory of trace dynamics, and also to Connes' non-commutative geometry (this latter for including gravity into trace dynamics, as a matrix dynamics). This is how we constructed an appropriate trace dynamics Lagrangian for gravity coupled to fermions, ensuring that quantum theory is emergent, and that it has the desired classical limit \cite{maithresh2019}. We then showed how Yang-Mills fields can be brought into the Lagrangian, in the conventional spirit of modifying the Dirac operator \cite{MPSingh}. Only subsequently it was realised that a fundamental explanation of spin \cite{Singhspin1} after including Yang-Mills interactions strongly indicates the doubling of space-time dimensions from four to eight. This is how octonions were implicated in our theory. The successful merging of two apparently disparate investigations [division algebras and standard model, versus trace dynamics with gravity]  strongly suggests that both the investigations are on the right track. The two investigations complement each other.

This is how the terms $T_4, T_5, T_6, T_7$ and $T_8$ in our Lagrangian relate to the algebra of complex octonions, and to the electro-color symmetry of the standard model. We now show that another copy of $C\ell(6)$ relates the terms $(T_1, T_2, T_3)$ in the Lagrangian to the other two interactions, gravitation and the weak force, via the gravito-weak symmetry, which we newly propose in this work. It turns out that these three terms will force us to extend the algebra to sedenions, that being the only way to include gravity in the standard model. But as a bonus, we will get three fermion generations. The existence of three generations is an inevitable consequence of unifying gravity with the standard model.

\bigskip

\subsubsection{Introducing the gravito-weak symmetry}
\bigskip
We start by summarising our new findings and then we justify them in  detail.

We have seen above that the symmetry group within  $G_2$, which describes the electro-color symmetry and the fermion properties, is $SU(3)$. It is the element-wise stabiliser group of octonions.  It turns out that the symmetry group 
which describes the first three terms $(T_1, T_2, T_3)$ of our Lagrangian is ${\rm Stab}_{G_2} (\mathbb H)$, the stabiliser group of the quaternions inside the octonions. This group happens to be $SO(4)$. It is the other maximal sub-group of $G_2$, beisdes $SU(3)$. This is the group which describes the gravito-weak symmetry. It is the group generated by the Clifford algebra $C\ell(6)$ which Stoica \cite{Stoica} constructs to describe the weak and Lorentz sector. A sub-group of the  stabiliser group ${\rm Stab}_{G_2} (\mathbb H)$ (this being  $SO(4)$),  is the element-wise stabiliser group of quaternions,
$\rm{Fix}_{G_2}(\mathbb H)$, which happens to be $SU(2)$. This $SU(2)$ describes the weak symmetry, corresponding to a $C\ell(4)$ Clifford algebra. Thus, just as the element-wise stabiliser group $SU(3)$ of octonions  describes the electro-colour symmetry, the element-wise stabiliser group $SU(2)$ of quaternions describes the weak symmetry. The groups $SU(3)$ and $SO(4)$ thus constructed have an intersection which is a $U(2)$ group. The $SU(2)$ of weak symmetry  is the simple part of $U(2)$ and is also a normal sub-group of $SO(4)$. In this manner, the weak symmetry is a part of both the standard model symmetries and the space-time gravito-weak symmetry. The $SO(4)$ is a group extension of $SO(3)$ -  the automorphism group of quaternions (i.e. $SO(3)$) - by the $SU(2)$.  We have seen that complex quaternions generate Lorentz symmetry. Thus, $SU(2)$  works as a bridge to unify the standard model with the Lorentz symmetry and thereby with gravity.  Also, the gravito-weak symmetry correctly maps the projector of $S^u$ basis (see previous section) to the particles in $S^d$. It explains why weak interactions violate parity. The two Clifford algebras $C\ell(6)$ constructed by Stoica are not completely different - they happen to have the $SU(2)$ weak symmetry overlap, in the group theoretic sense just mentioned. Thus effectively, the Clifford algebra they amount to is $C\ell(8)$. This then is the Clifford algebra which describes the unification of the standard model with Lorentz symmetry. However, $C\ell(8)$ cannot be constructed from the algebra of octonions. Therefore, following Gillard and Gresnigt \cite{Gillard_2019}, we are compelled to go beyond division algebras, onto complex sedenions. Because the minimal left ideals that the complex sedenions  generate leads to a useful $C\ell(7)$. Moreover, the automorphism group of sedenions is essentially three copies of the automorphism group $G_2$ of octonions, suggesting a way to get the three fermion generations \cite{Gillard_2019}. Remarkably, these three copies of $G_2$ have an intersection, which happens precisely to be the stabiliser group of the quaternions! Thus the gravito-weak symmetry is shared amongst the three generations, only the electro-colour part differs. The theory has four extra terms yet unaccounted for, which could describe the Higgs bosons. We predict a new spin one boson, the Lorentz boson [so named recently by Cahill \cite{Cahill:2020lry}], which describes the quantisation of the Lorentz symmetry. This, and not the graviton, is the gravitational analog of the photon.  After the universe undergoes spontaneous localisation and classical space-time emerges, the electro-weak symmetry becomes part of the internal symmetries of the standard model. Gravitation emerges essentially as a gauging of the Lorentz symmetry possessed by the individual aikyons. We note that above we have described only one aikyon, and the various bosons and leptons are its different manifestations.

We now explain in detail as to how we came to these conclusions. The group theoretic properties of $G_2$ mentioned above can be found, amongst other places, in the nice summary available at \href{https://ncatlab.org/nlab/show/G2\#Miyaoka93}{this\ link}, along with references. The cartoon below attempts to describe the relative role of the various symmetry groups.
\begin{figure}[!htb]
        \center{\includegraphics[width=\textwidth]
        {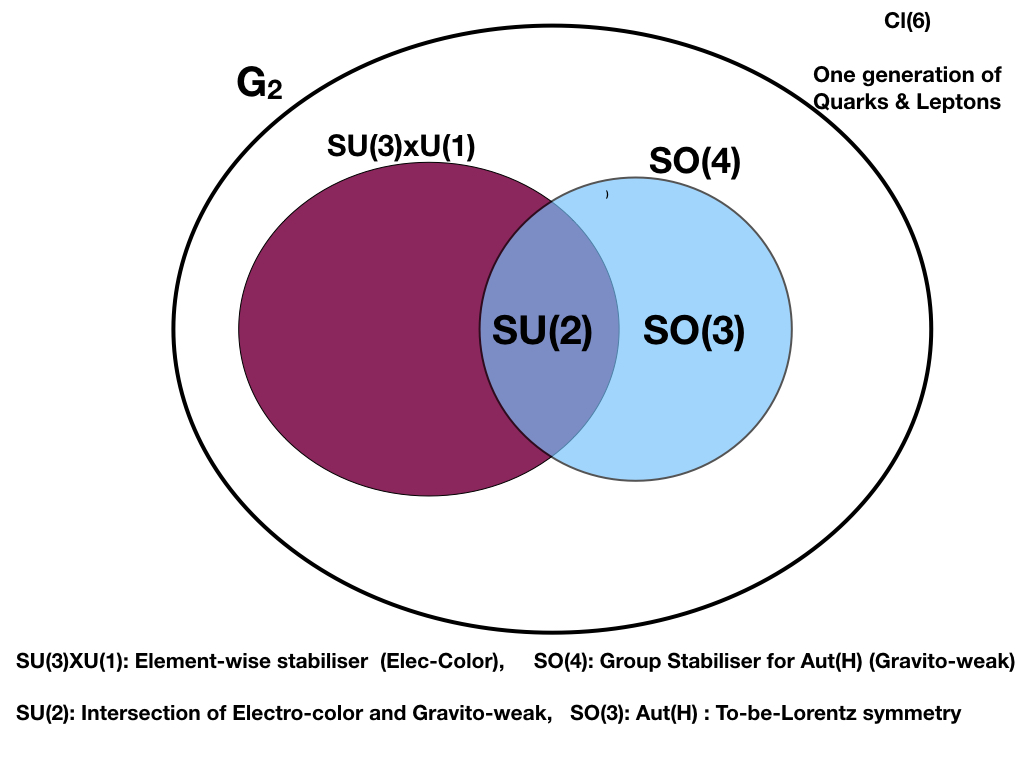}
        \caption{\label{fig:my-label3}} {The two maximal sub-groups of $G_2$ whose intersection is $U(2)\sim SU(2)\times U(1)$. This diagram shows the unification of the standard model with Lorentz symmetry and hence with gravity. The element stabiliser group of $G_2$ generates the electro-color symmetry $SU(3)$. The group stabilizer of ${\rm Aut} (\mathbb H)$ generates the gravito-weak symmetry $SO(4)\sim SU(2)\times SU(2)$. Their intersection is the weak symmetry $SU(2)$ enhanced by $U(1)_{em}$.  Gravito-weak extends $SO(3)$  by the weak symmetry because $SO(4)$ is a group extension of $SO(3)$ by $SU(2)$. Lorentz symmetry $SL(2,C)$ is constructed from generators of $C\ell(2)$ made from  ${\mathbb C}\times {\mathbb H}$ [complexification of $SO(3$), i.e. of ${\rm Aut(}{\mathbb H})$.]   }}
      \end{figure}
We rely heavily on the three recent and important papers by Furey \cite{f2}, by Gillard and Gresnigt \cite{Gillard_2019}, and by Stoica \cite{Stoica}. Incidentally, all these three papers came out as recently as 2019, and without these papers the present work would be impossible [`standing on the shoulders of giants']. 

Stoica constructs two copies of $C\ell(6)$ [no division algebras involved; only Clifford algebras] one of which describes the electro-colour symmetry, and matches precisely with the $C\ell(6)$ which Furey constructs and which we described above. The more interesting part is the second copy of $C\ell(6)$ constructed by Stoica. Four of the six generators for the Clifford algebra properly describe the action of $SU(2)$ weak symmetry on the quarks and leptons. The strange part is this. While these generators are not linear combinations of the generators of the other $C\ell(6)$, and in that sense not dependent on them, nonetheless they bear an intricate mathematical  relation to the electro-color generators. This is extremely surprising and suggestive. It is telling that the electro-colour part somehow knows about the weak symmetry [as Furey notes too, emphatically]. Moreover, while Stoica correctly adds the Lorentz symmetry to the $SU(2)$ to make the second copy of $C\ell(6)$, it is discomforting that Lorentz symmetry has to be added on by hand. For me this was a  strong signal that the weak and Lorentz symmetries must be unified in a fundamental way, using division algebras. And above all, the terms $T_1, T_2, T_3$ in the Lagrangian are begging for Lorentz-weak unification - this is  evident from their composition of these terms, their corresponding counter-part having  been taken up by electro-colour.

Furey also constructs a $C\ell(4)$, from the generators of the electro-colour $C\ell(6)$, and these correctly describe the right action of $SU(2)_{weak}$ on quarks and leptons. Let us recall how this was done. Given the nilpotent $\omega = \alpha_1 \alpha_2 \alpha_3$ made from the electro-colour generators of the previous section, the following $C\ell(4)$ generators are constructed from them:
$$(\tau_1 i\epsilon_1, \tau_2 i\epsilon_1, \tau_3 i \epsilon_1, i\epsilon_2),
\label{magic}
$$
where $(\epsilon_0, \epsilon_1, \epsilon_2, \epsilon_3)$ are the basis vectors of a quaternion, and where $\tau_1 \equiv \omega + \omega^\dagger, \tau_2 \equiv i\omega - i\omega^\dagger, \tau_3 \equiv \omega \omega^\dagger - \omega^\dagger \omega$. These generators are then written in a new basis $[\beta_1, \beta_2,  \beta_1^\ddag, \beta_2^\ddag]$ and are fermionic ladder operators, where
\begin{equation}
\beta_1 \equiv (i\epsilon_2 + i \epsilon_1 \tau_3); \qquad \beta_2 \equiv \omega^\dag i \epsilon_1
\label{iden}
\end{equation}
[These ladder operators are analogous to the $(\omega_u, \omega_u^\dagger, \omega_d, \omega_d^\dagger)$ constructed by Stoica \cite{Stoica}, but not the same ones].
Here, $\ddag$ stands for simultaneous complex conjugation, quaternion conjugation and octonion conjugation. These generators can be shown to form the Clifford algebra $C\ell(4)$.
An idempotent is constructed, and the ladder operators and minimal right ideals transform the leptons as expected for weak interactions, under an $SU(2)$ symmetry constructed from the following three $SU(2)$ generators:
\begin{equation}
T_1\equiv \tau_1 (1+ i\epsilon_3)   ; \qquad T_2 \equiv \tau_2 (1+ i\epsilon_3) ; \qquad T_3 \tau_3 \equiv \tau_3 (1+ i\epsilon_3)
\end{equation}
where $\epsilon_3$ is the third quaternion component, which was not used in making the $\beta$ generators. The chirality property of the leptons is automatically recovered under the application of this $SU(2)$. 

Let us look at this remarkable construction more closely, as it will guide us to the stabiliser group $SO(4) \sim SU(2)\times SU(2)$ of quaternions and the proposed gravito-weak symmetry. The $\beta$ generators are made from $G_2$ automorphisms acting on the $\alpha$ generators of the electro-colour symmetry, and from two of the three vectors of a quaternion. Let us {\it assume} that this quaternion belongs to the quaternion sub-algebra of the very octonions we are studying and which describe the standard model (i.e. $\epsilon\in e$). What are the implications of this assumption?  These $\beta$ generators result from automorphisms which belong to the $SU(3)$ group shown above in Fig. 2. Furthermore thery are made from acting on {\it only two} of the three quaternionic elements of the said quaternion. For the moment, just to indicate where we are headed, let us assume that $\tau_1, \tau_2$ and $\tau_3$ are different from the quaternionic components $\epsilon_1$ and $\epsilon_2$. Because $\epsilon_3$ is not transformed by the $\beta$ automorphisms, these automorphisms belong to the element-preserving group $SU(2)$ of the quaternions. [Just as $SU(3)$ is the element-preserving group of the octonions]. This appears to be the reason why the above construction works. So we can now look beyond this specific construction, and propose the following. Consider the action on quaternions, of those automorphisms inside 
the chosen $SU(3)$, which belong to the element-preserving group $SU(2)$ of quaternions. Ladder operators  made from these automorphisms  will transform under the  $(\tau_1, \tau_2, \tau_3)$ (which are used to construct the  $SU(2)$ generators), precisely as under an $SU(2)$ symmetry. We propose to identify this with the weak symmetry of the standard model. Weak interactions are that part of the electro-colour symmetry which belong to the element-preserving group $SU(2)$ of the quaternions inside the element-preserving group $SU(3)$ of the octonions. The physical reason behind this mathematical proposal remains to be understood, just as we do not know why electro-colour symmetry is described by the element preserving group of the octonions. The explanation will possibly come from the unified theory, which we will describe below, shortly.

Our  claim finds strong support in the recent research of other workers, on the maximal sub-groups of compact exceptional Lie-groups, and their relevance for the standard model. These developments are based on the Borel - di Siebenthal \cite{BdS} theory for classification of such groups. See especially the works of Todorov and Drenska \cite{Todorov:2018yvi}, Todorov and Dubois-Violette \cite{tkey}, and Baez and Huerta \cite{Baez}. Section 2.1 of the first of these papers has an elegant proof that the element-preserving sub-group inside $G_2$ is $SU(3)$ and the stabiliser group of the quaternions inside the octonions is $SU(2)\times SU(2)/{\mathbb Z_2}$. Eqn. (4.2) of the paper by Todorov and Dubois-Violette \cite{tkey} notes that the intersection of these two groups is $U(2)$, which happens to be the gauge-group for the Weinberg-Salam model, and has the $SU(2)_L$ sub-group inside it. Thus, following our proposal in the previous paragraph, we re-state that this $U(2)$ should indeed be seen as a part of the electro-colour symmetry, somehow suggesting that electro-weak follows from electro-colour. We will have more to say about this when we return to the terms $T_1, T_2, T_3$ in our Lagrangian. 

We now propose the gravito-weak symmetry as the automorphism-group of the other maximal-subgroup of $G_2$, namely $SU(2)\times SU(2)/{\mathbb Z}_2$. This group is also the group extension, by the afore-mentioned $SU(2)$, of the automorphism group ${\rm Aut} ({\mathbb H})$ of the quaternions. It is also the stabiliser group of quaternions, which means that automorphisms of $G_2$ belonging to this group, when acting on quaternions, will send them to other quaternions. This safety-net for the quaternions within the octonions is absolutely essential for the emergence of classical space-time [whose Lorentz symmetry group is ${\mathbb C}\times {\mathbb H}$; automorphisms of this group lead to general relativity, as we see below.]

We recall  our Lagrangian (\ref{fulllag2}) below, now without the adjointness condition imposed on $q_B$. Only $\dot{q}_{B}$ will be assumed to be self-adjoint.
 \begin{equation}
\begin{split}
 \frac{S}{C_0} = \frac{1}{2}\int \frac{d\tau}{\tau_{Pl}}\; Tr \bigg[\biggr.\frac{L_P^2}{L^2}\bigg\{\biggr.  \dot{q}_B^2  +\frac{L_P^2}{L^2} \dot{q}_B\beta_2 \dot{q}_F + \frac{L_P^2}{L^2} \beta_1 \dot{q}^\dagger_F \dot{q}_B   + \frac{L_P^4}{L^4} \beta_1\dot{q}^\dagger_F \beta_2\dot{q}_F  \\- \frac{\alpha^2 }{L^2} \bigg(q_B^\dagger q_B     
+ \frac{ L_P^2}{L^2} q_B \beta_2 q_F + \frac{ L_P^2}{L^2} \beta_1 q^\dagger_F q_B^\dagger +  \frac{L_P^4}{L^4} \beta_1 q^\dagger_F  \beta_2 q_F  \bigg)
 \biggl. \bigg\} \biggl.\bigg] 
 \end{split}
 \label{fulllag2rep}
 \end{equation} 
The sum of the first three terms is assumed to be invariant under automorphisms belonging to the stabiliser group of quaternions ${\rm Stab}_{G2}({\mathbb H})$. This is the gravito-weak symmetry. We recall that the bosonic indices are the quaternionic part of octonion indices and run from zero to three. The bosonic kinetic energy term (\ref{bofi}) is
\begin{equation} 
 \begin{split}
 \dot{q}_B^2  & = \bigg [  \dot{q}_{Be0} \; e_0 + \dot{q}_{Be1} \; e_1 + \dot{q}_{Be2}\ e_2 + \dot{q}_{Be4}\; e_4 \bigg ] \times \bigg [  \dot{q}_{Be0} \; e_0 + \dot{q}_{Be1} \; e_1 + \dot{q}_{Be2}\ e_2 + \dot{q}_{Be4}\; e_4 \bigg ] \\  & =  \dot{q}_{Be0}^2 - \dot{q}_{Be1}^2 -\dot{q}_{Be2}^2 -\dot{q}_{Be4}^2 + 2 \dot{q}_{Be0} \; \big[ \dot{q}_{Be1} \; e_1 + \dot{q}_{Be2}\ e_2 + \dot{q}_{Be4}\; e_4 \big]
 \label{bofi2}
 \end{split}
 \end{equation}
 
 Here it is evident that there must exist a  massless spin one `Lorentz' boson corresponding to this `quantisation/gauging' of the Lorentz symmetry. Such a particle has recently been proposed in a very interesting paper by Cahill in \cite{Cahill:2020lry} [the coinage Lorentz boson is due to him].  The discovery of such a boson would constitute a supportive evidence for our theory. Added to the nine bosons  coming from the electro-colour sector, the Lorentz boson is the tenth one. Now, $G_2$ has fourteen generators, so one needs four more bosons. On the other hand leaving out the $\dot q_{Be0}$ part we have three bosonic operators to account for. Thus it seems highly plausible that the $\dot{q}_B^2$ term - because ${\rm Aut}(\mathbb H)$ has the Clifford algebra $C\ell(2)$ and gives rise to a pair of generators - corresponds to two bosons, the Lorentz boson and the second one possibly the Higgs boson. Whereas the remaining three $\dot{q}_B$  components correspond to the three weak isospin bosons. How two of them acquire charge remains to be understood. Because electric charge is being fixed in the electro-colour sector. Undoubtedly, gravito-weak and electro-colour interact non-trivially in the intersection. As Furey has noted, the nilpotents $\omega$ and $\omega^\dagger$ constructed from the electro-colour ladder operators have electric charge, and also map isospin up to isospin down correctly when acting to the right on the idempotent.

Can one construct a $C\ell(6)$ algebra for the gravito-weak sector, which could be said to be independent from the electro-colour $C\ell(6)$? Without making additional assumptions / approximations, no.  Under a certain approximation, yes. The approximation would consist of first making the $SU(2)$ $\beta$ generators from electro-colour ladder operators, and then setting the electro-colour coupling constant to zero, which physically means that gravito-weak is a good effective symmetry under circumstances where strong and electromagnetic effects are insignificant. [We discuss the implications of this below.] The $\beta$ generators were made using four octonionic directions from outside the quaternion, and two more from inside the quaternion $(e_0, e_1, e_2, e_4)$. That leaves two directions for making a $C\ell(2)$ to describe Lorentz symmetry, if we suitably redefine and  make a quaternionic triplet from these two un-used directions. If we say label these directions as $e_1$ and $e_4$ then we can make a $C\ell(6)$ algebra from the following six ladder operators:
\begin{equation}
\begin{split}
\alpha_0 = \frac{1}{2} (ie_1 - e_2) ; \qquad \beta_1 = (ie_4 + i e_1 \tau_3); \qquad \beta_2 =  \omega^\dag i e_1 \\
 \alpha_0^\dagger =\frac{1}{2} (ie_1 + e_2); \qquad  \beta_1^\dagger = (ie_4 - i e_1 \tau_3);    ; \qquad \beta_2^\dagger =  ie_1 \omega 
 \end{split}
\end{equation}
This Clifford algebra implies a non-trivial mixing between the weak sector and the Lorentz sector, because only one of the bosons coming from the ${\rm Aut}({\mathbb H})$ sector has to do with gravity.  Hence a connection between the gravitational force and the weak force is implied, whose experimental consequences must be explored. To our understanding, these ladder operators are different from the ones constructed by Stoica \cite{Stoica}.

We also note another promising avenue for studying the gravito-weak symmetry, which could come from investigating the action of complex quaternions on octonions. We have recently studied the trace dynamics of what we called pure gravity case \cite{maithresh2019} [no internal symmetries]. At that time we expected weak interaction to also be part of an internal symmetry - a Yang-Mills field like the other two [electro-colour]. But now we know that the weak-interaction is a short-range space-time symmetry, but not pertaining to 4-D spacetime; it pertains to the 8-D octonionic spacetime. Also, note that there is no dimensionless coupling constant $\alpha$ for the weak interaction, neither in our theory nor in the standard model. [Strong interactions and electrodynamics of course have a dimensionless coupling constant]. The weak coupling constant, i.e. the Fermi constant $G_F$, is dimensionful, like Newton's $G_N$ for gravity. In fact, we will show below that our theory gives the correct value of $G_F$, from $G_N$, if we make use of the known value of mass for the Higgs. 

 The Lagrangian which we studied, thinking of it as the pure gravity case \cite{maithresh2019}, and which we now re-write here allowing an adjoint of $\dot{q}_F$ to be taken, is
 \begin{equation}
\begin{split}
\frac{S}{C_0}  =  \frac{1}{2} \int \frac{d\tau}{\tau_{Pl}} \; Tr \bigg[\frac{L_P^2}{L^2c^2}\; \left(\dot{q}_B +\beta_1 \frac{L_P^2}{L^2}\dot{q}^\dagger_F\right)\; \left(\dot{q}_B +\beta_2 \frac{L_P^2}{L^2} \dot{q}_F\right) \bigg] = \\
 \frac{1}{2}\int \frac{d\tau}{\tau_{Pl}}\; Tr \bigg[\biggr.\frac{L_P^2}{L^2}\bigg\{\biggr.  \dot{q}_B^2  +\frac{L_P^2}{L^2} \dot{q}_B\beta_2 \dot{q}_F + \frac{L_P^2}{L^2} \beta_1 \dot{q}^\dagger_F \dot{q}_B   + \frac{L_P^4}{L^4} \beta_1\dot{q}^\dagger_F \beta_2\dot{q}_F 
 \biggl. \bigg\} \biggl.\bigg] 
 \end{split}
 \label{gravweak}
 \end{equation} 
This is precisely the same Lagrangian as our present Lagrangian (\ref{fulllag2rep}) above, but now with the coupling constant $\alpha=0$. Knowing that the symmetry group for this Lagrangian is the other maximal sub-group $SU(2)\times SU(2)/{\mathbb Z}_2$ of $G_2$, and noting that $\dot{q}_B$ is a quaternion, we can construct a $C\ell(6)$ algebra from the left minimal ideals of complex quaternions acting on complex octonions. This will give the correct right action on the eight fermion basis $S^u$ and again constitutes evidence for the gravito-weak symmetry. It is mediated by five bosons - the three weak isospin bosons, the Lorentz boson, and another which possibly plays the role of the Higgs boson. We plan to study this Lagrangian in detail, from the division algebra viewpoint, in a forthcoming investigation.

The idea of a gravi-weak unification in the context of quantum field theoretic approaches has been investigated also by Nesti and Percacci \cite{Percacci}. They note that the complexified $SO(3,1)$ has two copies of the $SU(2)$ sub-algebra, one of which can be identified with the weak interaction, and the other with gravity. We have come to essentially the same conclusion in our analysis above, from quantum foundational considerations.

When the electro-colour and gravito-weak sector are both to be taken into account, the two Clifford algebras are not independent but have a $C\ell(4)$ overlap between them. Thus the algebra is effectively a $C\ell(8)$ algebra, which cannot be made from complex octonions. This compels us to consider the next algebra, beyond the octonions,  in the Cayley-Dickson construction, namely the sedenions. However, we pursue this promising path only because because the aikyon can still be an octonion, with the complex sedenions effectively yielding three copies of the octonion algebra. This is very promising, as it can explain the three fermion  generations, besides the sedenions providing a promising path towards unifying gravity with the standard model interactions. Once again, we will construct left minimal ideals from the action of complex sedenions onto themselves. We almost wholly follow the recent important work of Gillard and Gresnigt \cite{Gillard_2019}, interpreting  it in the context of our Lagrangian.

A second promising avenue, which also fits our Lagrangian very well, is the extension beyond division algebras, to the Jordan algebra $J_3({\mathbb O})$ of 3x3 Hermitean matrices with octonionic entries. This builds on the important and beautiful recent work of Dubois-Violette, Todorov, and their collaborators, on exceptional Jordan algebras, especially concerning the automorphism group $F_4$, of the algebra of $J_3({\mathbb O})$, of which $G_2$ is a sub-group. We discuss both the Jordan algebra approach and the sedenion approach applied to  our Lagrangian. It appears at this juncture that the Jordan algebra approach is suited to the self-adjoint part of our Lagrangian. Whereas the sedenion approach seems more relevant for the full Lagrangian, which is not self-adjoint. The anti-self-adjoint part is essential for recovery of the classical limit via spontaneous localisation. We emphasise again that this Lagrangian was not constructed to suit the investigations of division algebras, Jordan algebras, and sedenions. It was constructed to arrive at a formulation of quantum field theory which does not depend on classical time, using the methods of trace dynamics  and Connes'  non-commutative geometry. Hence it is very encouraging that the Lagrangian is invariant under automorphisms induced by a suitable algebra and describes well the standard model and its unification with gravity.

\bigskip

\subsubsection{Towards unification}
We now work with the full Lagrangian of our theory, in which the total time derivative terms were not dropped, and which is given by [Eqn. (11) of \cite{MPSingh}]
\s[
\mathcal{L} = Tr \biggl[\biggr. \dfrac{L_{p}^{2}}{L^{4}} \biggl\{\biggr. i\alpha \biggl(\biggr. q_{B}^\dagger + \dfrac{L_{p}^{2}}{L^{2}}\beta_{1} q_{F}^\dagger \biggl.\biggr) + L \biggl(\biggr. \dot{q}_{B} + \dfrac{L_{p}^{2}}{L^{2}}\beta_{1}\dot{q}_{F}^\dagger \biggl.\biggr) \biggl.\biggr\}\\
\biggl\{\biggr. i\alpha \biggl(\biggr. q_{B} + \dfrac{L_{p}^{2}}{L^{2}}\beta_{2} q_{F} \biggl.\biggr) + L \biggl(\biggr. \dot{q}_{B} + \dfrac{L_{p}^{2}}{L^{2}}\beta_{2} \dot{q}_{F} \biggl.\biggr) \biggl.\biggr\} \biggl.\biggr]
\label{wowlag}
\s]
The Lagrangian (\ref{fulllag2rep}) we have been using so far is a special case of this complete Lagrangian from which the total time-derivative terms have been dropped. We are going to need this full Lagrangian if we are to describe three generations of fermions - only then there are enough degrees of freedom. In fact the count of number of terms matches perfectly with what is desired. The Lagrangian (\ref{fulllag2rep}) has $128$ terms [$16\times 8$]. The full Lagrangian above has $256$ terms. We are going to demand that the gravito-weak sector [64 terms] should be common amongst the three generations [because it describes space-time symmetry, not matter content]. The electro-colour sector of each generation requires 64 terms, and for three generations this comes to $64\times 3 = 192$ terms. Add to that  the $64$ of gravito-weak part and we get  $256$. This is indeed remarkable - it matches perfectly with  the number of terms in the full Lagrangian! We find this highly encouraging. 

Next, we note that this full Lagrangian is a product of two terms each of which is the sum of two octonions (only one of them has the real direction in it).  This strongly suggests that we can consider each bracket as a complex sedenion, and the Lagrangian is a product of two sedenions. This motivates us to relate to the recent work of Gillard and Gresnigt, and obtain the unification of the standard model [three generations] with gravity, through minimal left ideals made from the action of complex sedenions on themselves. We outline this construction below, as to what we anticipate [which is somewhat different from Gillard and Gresnigt's conclusions], leaving the detailed analysis for future work.
The automorphism group of the  sedenions is
\begin{equation}
{\rm Aut}({\mathbb S}) = {\rm Aut} ({\mathbb O}) \times S_3
\end{equation}
where $S_3$ is the permutation group. It is isomorphic to Spin(8), famous for its triality. This suggests that we could think of these automorphisms as being made from three copies of $G_2$ automorphisms, two copies at a time. Such an inference then supports the idea that the three fermion generations arise for this reason, and this inference also dictates how we interpret the Clifford algebra made from sedenion automorphisms. The relation between three copies of Spin(8) and $G_2$ is elaborated at, for instance this link: {$https://ncatlab.org/nlab/show/SO\%288\%29$}  The exceptional group $G_2$ is the intersection of any two of the three Spin(7) sub-groups of Spin(8). Thus there are three such intersections and three copies of $G_2$ arise. Each copy of $G_2$ has a sub-group $SU(3)$ which has a sub-group $SU(2)$. We have also seen above that the $SU(2)$ is the group extension of the ${\rm Aut}({\mathbb H})$ inside $G_2$. Put together, these features suggest the pattern of symmetry groups for three generations as shown in Figure 3. The Lorentz symmetry is the three-way intersection amongst the three generations. Then there are three copies of pair-wise interactions between any two generations, mediated by electroweak interactions. Each generation has its own $SU(3)_c$ symmetry. There are eight gluons, the photon, three weak bosons, four possible Higgs bosons [see next sub-section], {\it and} the newly proposed Lorentz boson.
\begin{figure}[!htb]
        \center{\includegraphics[width=\textwidth]
        {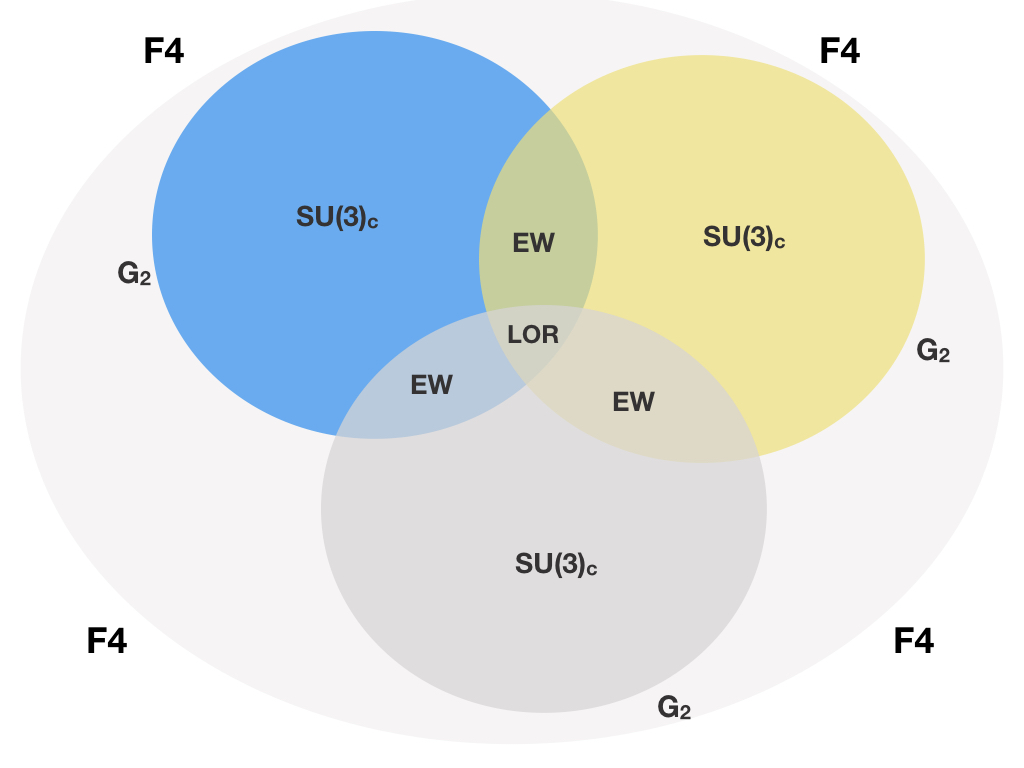}
        \caption{\label{fig:my-label6}} {The proposed unification of the three fermion generations of the standard model, and the unification of the standard model interactions with the Lorentz interaction and hence with gravity. The symmetry group of each of the three generations is $G_2$ and the three copies of $G_2$ are all embedded in  one copy of $F_4$ }.  
        The $G_2$ have a three-way intersection which is the Lorentz symmetry, and they have three pairwise intersections which is the electro-weak symmetry. Each fermion generation has its own $SU(3)_c$ symmetry. The unification symmetry group is $F_4$, the automorphism group of the exceptional Jordan algebra.}
      \end{figure}
One can construct a $C\ell(7)$ Clifford algebra from the fourteen out of the fifteen imaginary one-vectors of the sedenions. There are fourteen ladder operators which when acting on the idempotent, using products of zero to seven ladder operators at a time, generate a 128 dimensional basis, the analog of the $S^u$, ($128 = {}^7C_0     + {}^7C_1 + {}^7C_2 + {}^7C_3 + {}^7C_4 + {}^7C_5 + {}^7C_6 + {}^7C_7)$. Subject to confirmation by further investigation, the following is how we expect these 128 operator products to act on the three fermion generations. For each generation, 32 operators act to map the fermions onto themselves or each other within the same generation. These are: eight for $SU(3)_c$, eight for the weak symmetry, eight for the newly proposed Lorentz boson, eight for the Higgs boson accompanying the Lorentz boson. That equals 32$\times$3=96 for three generations. Between every pair of generations, 8 actions, making it 24 for the three pairs of generations. That adds up to 120. The remaining 8 out of the 128 could possibly come from the three potential Higgs boson like terms [that arose from terms $T_4$ and $T_8$ in the Lagrangian. Between the three of them they will create another eight operators which could act on the eight fermions in a degenerate manner, without making a distinction across the three generations. See also the work of Furey on three fermion generations \cite{f2}.

Under suitable circumstances, to be described below, spontaneous localisation will separate the gravito-weak symmetry from electro-colour, leading to the emergence of classical space-time. Perhaps the correct description of the full symmetry is as gravito-electro-weak-colour symmetry. It being understood that the weak symmetry is a spacetime symmetry as well as an internal symmetry. It modifies gravity on small scales.

This is as far as we are able to go with the construction of the Clifford algebra for the unified theory, at present. We believe we have presented a promising theory for the unification of interactions. Now, it turns out that the  full Lagrangian  (\ref{wowlag}) is related to the exceptional Jordan algebra $J_3({\mathbb O})$ in a very interesting way. This will likely help us predict the values of the standard model parameters at the relatively low energies at which accelerator experiments are currently being carried out.
Fig. 4 below depicts this unification.
\newpage
\begin{figure}[H]
        \center{\includegraphics[width=\textwidth]
        {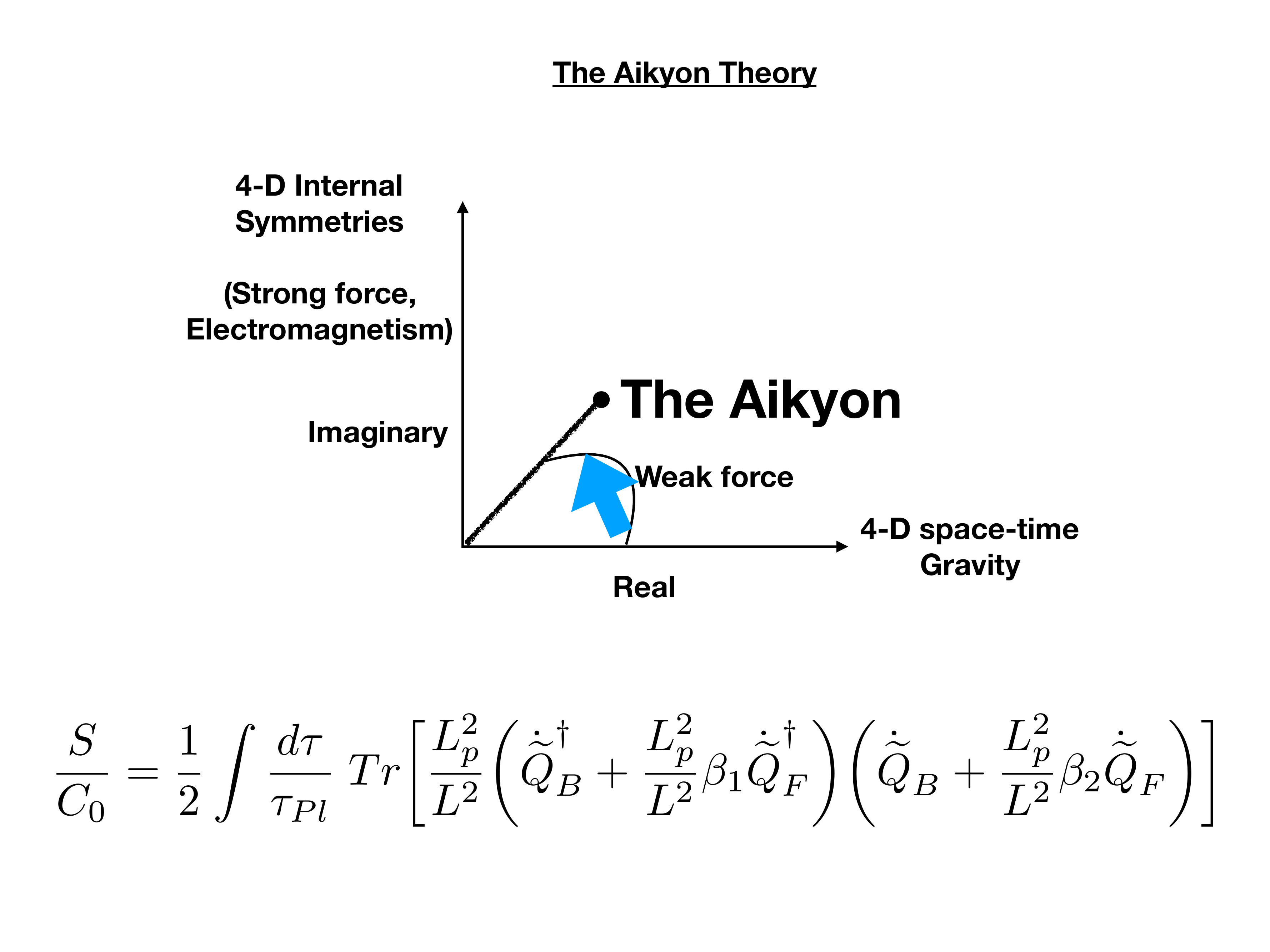}
        \caption{\label{fig:my-label5}} }\begin{singlespace}{ The Aikyon Theory: At the Planck scale, there is no distinction between space-time symmetry and internal symmetry. Physical space is eight dimensional non-commutative octonionic space. One can imagine it as a 2-D complex plane, where the real axis represents 4-D to-be-spacetime, and the imaginary axis represents 4-D to be internal symmetries. The aikyon is an elementary particle, say an electron, *along with* the fields it produces. We do not make a distinction  between the particle and the fields it produces. This is evident from the form of the action for an aikyon, shown above: variables with subscript B stand for the four known forces, and those with subscript F for any of the 24 known fermions of the three generations of the standard model. The Lagrangian is unchanged if B and F variables are interchanged. This is super-symmetry. And since the B-variables include both gravity and gauge-fields, there is a gauge-gravity duality.
The aikyon evolves in this 8-D space in Connes time. The aikyon is a 2D object, as if a membrane [2-brane]. Motion along the real axis is caused by gravity, along vertical axis by electro-colour force, and from real to imaginary by the weak force. Or we can just say, the aikyon moves in the 8D space under the influence of the unified force, given by the B-variable in the action.  There is one such action term for every aikyon in this space. Different aikyons interact by `colliding' with each other. The coordinates of this 8D space are the eight components of an octonion. Algebra automorphisms transform one coordinate system to another. These are the analog of general coordinate transformations of general relativity and internal gauge symmetries of gauge theories, and hence unify those concepts. The theory is invariant under 8D algebra automorphisms. And because the laws of motion are those of trace dynamics, this is already a quantum theory.
} \end{singlespace}
 \end{figure}

\bigskip

\subsubsection{Octonionic quantum mechanics, and the exceptional Jordan algebra   $J_3({\mathbb O})$}
The exceptional Jordan algebra is the algebra of 3$\times$ 3 Hermitean matrices with octonionic entries, closed under the Jordan product. An element $X(\xi, x)$ of the algebra is the matrix
\begin{equation}
X(\xi, x)=
\begin{bmatrix}
\xi_1 &  x_3 & x_2^*\\
x_3^* & \xi_2 & x_1\\
x_2 & x_1^* & \xi_3
\end{bmatrix}
\end{equation}
The product rules are given for instance in the paper of Dubois-Violette and Todorov \cite{tkey}. What is of great importance for us is the third order characteristic equation for $X(\xi, x)$ given by
\begin{equation}
X^3 - Tr (X) X^2 + S(X) X - det(X) = 0; \qquad Tr(X) = \xi_1 + \xi_2 + \xi_3 \qquad 
\end{equation}
where the determinant is
\begin{equation}
det (X) = \xi_1 \xi_2 \xi_3 + 2 Re (x_1 x_2 x_3) - \sum_1^3 \xi_i x_i x_i^*
\end{equation} 
and where $S(X)$ is given by
\begin{equation}
S(X) = \xi_1 \xi_2 - x_3 x_3^* + \xi_2 \xi_3 - x_1 x_1^* + \xi_1 \xi_3 - x_2^* x_2
\end{equation}
Let us compare this form of $S(X)$ with the operator Lagrangian (\ref{wowlag}) which we re-write as
\begin{equation}
 {\cal L} = Tr \bigg[ \frac{L_P^2}{L^2} \bigg\{ O_1 + O_3 \bigg\}  \bigg\{O_2 + O_4\bigg\}\bigg]
\end{equation}
where the octonions $O_1, O_2, O_3, O_4$ are defined as
\begin{equation}
\begin{split}
O_1 =L \biggl(\biggr. \dot{q}_{B} + \dfrac{L_{p}^{2}}{L^{2}}\beta_{1}\dot{q}_{F}^\dagger \biggl.\biggr) \biggl.\biggr\} ; \qquad O_2 =L \biggl(\biggr. \dot{q}_{B} + \dfrac{L_{p}^{2}}{L^{2}}\beta_{2} \dot{q}_{F} \biggl.\biggr) \biggl.\biggr\} \\
O_3 = i\alpha \biggl(\biggr. q_{B}^\dagger + \dfrac{L_{p}^{2}}{L^{2}}\beta_{1} q_{F}^\dagger \biggl.\biggr); \qquad O_4 = \biggl\{\biggr. i\alpha \biggl(\biggr. q_{B} + \dfrac{L_{p}^{2}}{L^{2}}\beta_{2} q_{F} \biggl.\biggr)
\end{split}
\end{equation}
The octonionic product inside the trace can be expanded out, and 
\begin{equation}
O \equiv  O_1 O_2    +   O_1 O_4 + O_3 O_2 + O_3 O_4
\end{equation}
On the face of it, it seems that we have four octonions here, not six, and that we are dealing with $J_2(8)$, not $J_3(8)$. It is known that $J_2(8)$ can describe one generation of standard model fermions. However, we recall that the Lagrangian we are dealing with is not self-adjoint. The relevant Lagrangian for $J_3(8)$ will be its self-adjoint part, which in any case is what survives in the emergent theory, and this self-adjoint part is the correct one for studying octonionic quantum mechanics. [Note however that unification is described by the full Lagrangian and not by its self-adjoint part]. Hence, writing each octonion as a sum of its self-adjoint part and its anti-self-adjoint part, and retaining only the self adjoint part of the sum $O$, we get
\begin{equation}
O_s = O_{1s}O_{2s} + O_{1as}O_{2as}  + O_{1s}O_{4s} + O_{1as} O_{4as} + O_{3s}O_{2s} + O_{3as}O_{2as} + O_{3s} O_{4s}  + O_{3as} O_{4as}
\end{equation}
Now we have eight octonions, and eight products, two more than desired for $J_3(8)$. However, it should be noted that the gravito-weak symmetry is common amongst the three generations, so there is likely some degeneracy in these terms. Furthermore, from our earlier work we know that this system has oscillatory solutions as a function of the Connes time $\tau$. So, on-shell, there are possibly some relations between these octonions. This issue is currently under investigation. For now, we will assume that the correct algebra for the automorphism invariance of the self-adjoint Lagrangian is the exceptional Jordan algebra $J_3({\mathbb O})$. The associated group of automorphisms is the exceptional Lie group $F_4$ which contains $G_2$  as a sub-group. In fact, because $F_4$ contains $Spin(8)$ as a sub-group, the triality property of $SO(8)$ leads to three copies of $G_2$ as described in the previous section. The symmetry structure described for the unified theory in Fig. 3 is in need of a larger group in which it is embedded, and that group is $F_4$. At this point we make contact with the work of Dubois-Violette, Todorov,  and their collaborators, who describe the particle physics predictions arising from $F_4$. We note here the importance of $SO(4)$, the stabiliser group of the quaternions in the octonions - this group is most essential for unifying the standard model with the Lorentz symmetry, and hence with gravitation. It is embedded in $G_2$. 

The characteristic equation will now depend on the coupling constant $\alpha$ and the length scale $L$, via the dimensionless constant $\alpha L_P/L$. Moreover, it is a cubic equation. This equation likely determines the masses of elementary particles of the three generations, and other standard model parameters. In our recent work \cite{q1q2uni}  we have argued that the ratio $\alpha L_P/L$ is independent of the energy scale. Hence at a length scale much larger than $L_P$, it should be possible to determine the running value of $\alpha$, after fixing a value for it at the Planck scale. This is a possible way to overcome the hierarchy problem - it is something natural to trace dynamics. Because in trace dynamics, which operates at the Planck scale,  the emergent quantum field theory is obtained by coarse-graining the underlying theory over length scales much larger than Planck length. As a result, the coupling constants of the emergent theory carry a memory of, and are determined by, their values at the Planck scale.

From the viewpoint of our theory, what is the connection between the complex sedenions, and the exceptional Jordan algebra? The complex sedenions are essential for constructing the Clifford algebra which describes the properties of the elementary particles. However, the automorphisms attached to the complex sedenions are embedded in $F_4$. This is also the automorphism group for the exceptional Jordan algebra: this algebra is essential for describing the octonionic quantum mechanics of the emergent quantum theory. Also, without this Jordan algebra, there will be no characteristic equation. 

As a consequence of our investigations in the present paper, we would like to propose the following. The unification of the standard model with gravitation is correctly described by the Lagrangian (\ref{wowlag}), in the framework of trace dynamics. The symmetry group which describes this unification is the exceptional Lie group $F_4$. The Clifford algebra describing the elementary particles is constructed from complex sedenions.  The symmetry of the self-adjoint part of the Lagrangian is described by the exceptional Jordan algebra 
$J_3({\mathbb O})$. The associated cubic characteristic equation determines the masses of the elementary particles, and the parameters of the standard model. Also, we predict the existence of the massless spin one Lorentz boson.

 \subsection{The Four Higgs bosons?} 
 The terms $T_4$ and $T_8$ in the Lagrangian, discussed in Eqn. (\ref{higgs}) above, have not been used up or interpreted so far. There will be one such pair of terms for each fermion generation and each pair likely constitutes one Higgs boson, which is used up to give mass to particles of that generation. In all probability, this mechanism, which remains to be worked out, works between a pair of generations, via electro-weak interaction, as suggested in Fig. 3 above. In addition, there is one additional boson generated along with the Lorentz boson, as we saw above, via the symmetry LOR  common across the three generations, as shown in Fig. 3. This could possibly be the Higgs boson observed in accelerator experiments. We hope to investigate the Higgs mechanism in our theory in the near future.
 
 In this  way, all the terms in our Lagrangian (\ref{wowlag})  have been accounted for. They describe the unification of three generations of standard model fermions and bosons, with the Lorentz interaction, and hence with gravitation. The symmetry group which describes this unification is $F_4$. 
 
 \subsection{Physical motivation for the gravito-weak unification}
 Unlike the electromagnetic interaction and the strong interaction, which have dimensionless coupling constants [to be determined from the coupling $\alpha$ in our Lagrangian], the weak interaction coupling constant $G_F$ is dimension-ful,  just as Newton's gravitational constant $G_N$ is. In fact, $G_F$ has dimensions of $(\hbar/c)^2\; G_N $. This suggests that the weak interaction is more like gravitation, and is a short-range [relic of] space-time symmetry, rather than an internal symmetry like electro-colour. In fact, at the Planck-scale, there is no distinction between internal symmetries and space-time symmetry. Only after spontaneous localisation separates space-time from the rest, the other symmetries are referred to as internal symmetries, and we club weak with electro-colour. Strictly, this isn't quite so. The weak interaction is the small-scale quantum counterpart of gravito-weak, and takes place in 8-D octonionic space, just like the electro-colour symmetry in the quantum domain acts in the 8-D space.
 
 Moreover, weak interactions violate parity, the only interaction which does so. And parity violation has to do with the relationship between spin and space-time, again suggesting a gravito-weak connection. We have recently given \cite{Singhspin} an explanation for the fundamental origin of spin. In our trace dynamics based Lagrangian, there ought to exist a definition of spin angular momentum, so that it is the canonical momentum corresponding to time-rate of change of some angle. And indeed that is how it turns out to be. When space-time is extended to the 8-D octonionic space, an angle can be constructed, so that it describes the evolution of a dynamical variable from the quaternionic direction [the horizontal `plane' defined above] to the imaginary octonionic directions that describe the `vertical' plane and where the electro-colour bosons lie. Spin is precisely the time rate of change of this angle (see Fig. 4 above), and the weak interaction is responsible for this motion from the four quaternionic directions to the other four directions of the octonion. It relates to the $SU(2)$ part of the maximal sub-group $SO(4)$ of $G_2$, also a group extension of ${\rm Aut}({\mathbb H})$. And now it is clear why SU(2) behaves as an internal symmetry as well as a space-time symmetry. This also provides a very clear understanding of why the weak interaction violates parity. The direction  of spin [and hence  handedness] with respect to the direction of motion in space-time, is determined by whether the departure from the horizontal plane is into the upper-half complex  plane [left-handed, anti-clock-wise, an $\exp i\theta$, with $\theta$ positive] or whether the departure is into the lower-half complex plane [right-handed / clockwise / an $\exp - i \theta$, with $\theta$ negative]. The change of direction of spin and hence of handedness can be described by a complex conjugation of the dynamical variable's evolution in the octonionic space. But complex conjugation also changes particles to anti-particles in our theory! It is no wonder then, that the weak interaction violates parity. It is in the very definition of spin and its handed-ness, and the definition of anti-particles as complex conjugates of particles. Hence fermions are necessarily chiral. We can as well think of anti-particle as a particle with its spin-flipped. We will also investigate if the CP-violation observed in the neutral kaon system can be attributed to the anti-self-adjoint term present in the Lagrangian in our theory.
 
 We can now relate $G_F$ and $G_N$ in a fairly straightforward way, from a knowledge of the mass of the Higgs boson. For this purpose we reproduce the gravito-weak Lagrangian Eqn. (\ref{gravweak}) below:
 \begin{equation}
\begin{split}
\frac{S}{C_0}  =  \frac{1}{2} \int \frac{d\tau}{\tau_{Pl}} \; Tr \bigg[\frac{L_P^2}{L^2}\; \left(\dot{q}_B +\beta_1 \frac{L_P^2}{L^2}\dot{q}^\dagger_F\right)\; \left(\dot{q}_B +\beta_2 \frac{L_P^2}{L^2} \dot{q}_F\right) \bigg] = \\
 \frac{1}{2}\int \frac{d\tau}{\tau_{Pl}}\; Tr \bigg[\biggr.\frac{L_P^2}{L^2}\bigg\{\biggr.  \dot{q}_B^2  +\frac{L_P^2}{L^2} \dot{q}_B\beta_2 \dot{q}_F + \frac{L_P^2}{L^2} \beta_1 \dot{q}^\dagger_F \dot{q}_B   + \frac{L_P^4}{L^4} \beta_1\dot{q}^\dagger_F \beta_2\dot{q}_F 
 \biggl. \bigg\} \biggl.\bigg] 
 \end{split}
 \label{gw2}
 \end{equation} 
 Let us recall that we have ($\dot{q}_B = \dot{q}_{Be0} e_0 + \dot{q}_{Be1} e_1 + ...)$. Let us substitute this in the Lagrangian, and also replace $C_0$ by its emergent value, i.e. $\hbar$, and bring it to the right side. We have
 \begin{equation}
 S \sim \int \frac{d\tau}{\tau_{Pl}} Tr \bigg[ \frac{\hbar L_P^2} {L^2 c^2}\bigg( \dot{q}_{Beo}^2 - \dot{q}_{Be1}^2 \bigg) + \frac{\hbar L_P^2}{L^2} \dot{q}_{Be0} \dot{q}_F + \frac{\hbar L_P^2}{L^2} \dot{q}_{Be1} \dot{q}_F  \bigg]
 \end{equation}
 where we have retained just enough structure [self-explanatory] that will suffice to show the relation between $G_N$ and $G_F$. In the emergent theory, after classical space-time arises, the trace is replaced, as a result of the heat-kernel expansion, by $\int \sqrt{g}\; d^4 x /L_P^4$. Making this substitution we get
 \begin{equation}
 S \sim  \int \frac{d\tau}{\tau_{Pl}} \int \sqrt{g} \; d^4x \bigg [ \frac{\hbar}{L_P^2 L^2 } \dot{q}_{Be0}^2   - \frac{\hbar}{L_P^2 L^2 } \dot{q}_{Be1}^2  + \frac{\hbar L_P^2}{L^2} \dot{q}_{Be0} \dot{q}_F + \frac{\hbar L_P^2}{L^2} \dot{q}_{Be1} \dot{q}_F \bigg]
 \end{equation}
The conventional Dirac operator $D_B$ is related in our theory to $\dot{q}_{Be0}$ by $D_B = (\hbar /L)\dot{q}_{Be0}$. The heat-kernel expansion also tells us that its square is proportional, in the classical limit, to the Ricci scalar. Moreover, $G_N$ is by definition, given by $L_P^2 = G_N\hbar/c^3$. Also, we had defined the fermionic part of the Dirac operator: $D_F \sim (1/L) \dot{q}_F$. Putting this in gives that,
\begin{equation}
 S \sim  \int \frac{d\tau}{\tau_{Pl}} \int \sqrt{g} \; d^4x \bigg [ \frac{c^3}{G} R   - \frac{\hbar }{L_P^2 L^2 } \dot{q}_{Be1}^2  + \frac{\hbar  }{L_P^2 } D_B D_F + \frac{\hbar}{L_P^2 L  } \dot{q}_{Be1} D_F \bigg]
\end{equation}
Further, it was argued   \cite{maithresh2019} that in the classical limit, spontaneous localisation localises $D_B$ to $1/L$, $D_F$ to $L_P^2/L^3$, and that $1/L^3$ can be replaced by the spatial $\delta$-function, since $L=\hbar /mc$ becomes the Compton wavelength, where $m$ is the mass of the particle. This gives
\begin{equation}
 S \sim  \int \frac{d\tau}{\tau_{Pl}} \int \sqrt{g} \; d^4x \bigg [ \frac{c^3}{G} R   - \frac{\hbar }{L_P^2 L^2 } \dot{q}_{Be1}^2 \bigg] + mc \int ds  + \frac{\hbar}{L} \int ds \dot{q}_{Be1}  
\end{equation}
Thus we have the gravitational action, the familiar action for the relativistic point particle, {\it and} terms for the weak interaction, which we can now interpret. We note that in the definition of $D_B$, the length scale $L$ was used, whereas $L_P$ stayed out. Now, in defining a potential $A_W$ for the weak interaction, we define $A_W \equiv \dot{q}_B/L_P$. 
And in analogy with the definition of $G_N$, we define $G_F$ by $\hbar c/G_F =  1/L^2$, giving $G_F /\hbar c= L^2 \implies G_F\sim  (\hbar/c)^2 (L/L_P)^2 G_N$. If we take $L$ to be the length scale  corresponding to the Higgs VEV $\sim$ 246 GeV, we get $G_F \sim (10^{33} \hbar /c)^2 G_N  \sim 10^{-62}$ MKS units \cite{Onofrio1}. In this way, $G_N$ determines $G_F$. Basically, $G_F/\hbar c$ is related to $G_N$ via the factor $L^2 / L_P^2$ suggesting a duality symmetry which ought to be investigated further. It is indeed highly suggestive that the relation between $G_F/\hbar c$ and $G_N$ can be written as
\begin{equation}
\frac{G_F/\hbar c}{L^2} = \frac{G_N/c^3\hbar}{L_P^2}
\end{equation} 
 
 Finally, we note that in the above Lagrangian, if we ignore $G_N$ and its coupling to mass, and pull the weak coupling constant $G_F$ out, so that it appears in front of the interaction term [the last term in the Lagrangian on the right above] we get that this interaction term is of the form $(G_F/\hbar c) (m/m_{Pl}) \int ds A_W $. This suggests a modification to the gravitational field in the microscopic regime.

The idea of a gravi-weak unification in the context of quantum field theoretic approaches has been investigated also by Nesti and Percacci \cite{Percacci}. They note that the complexified $SO(3,1)$ has two copies of the $SU(2)$ sub-algebra, one of which can be identified with the weak interaction, and the other with gravity. Strong physical motivation for relating gravity and the weak force arises also in the work of  Onofrio \cite{Onofrio1, Onofrio2}. 
 
 \section{Emergence of quantum theory below the Planck scale}
 We can now describe  the trace dynamics equations of motion without having to refer the matrices to the octonionic coordinate system.
 The equations of motion can be derived in a compact way from the Lagrangian (\ref{lagnew}) which we reproduce here for easy reference \cite{q1q2uni}
  \begin{equation}
Tr {\cal L} = \frac{1}{2} a_1 a_0\;   Tr \left[\dot{q}_1^\dagger  \dot{q}_2  - \frac{\alpha^2 c^2}{L^2} q_1^\dagger  q_2  \right]
\label{lagnew2}
\end{equation}
 where $S\equiv \int d\tau \; {Tr \cal L}$ and   $a_0 \equiv L_P^2 / L^2$ and $a_1 \equiv C_0 /cL_P $. Also, $q^\dagger_1 = q_B^\dagger + \beta_1 q^\dagger_F$ and $q_2 = q_B + \beta_2 q_F.$ 
 Variation of this Lagrangian with respect to $q_1^\dagger$ and $q_2$ gives the following two Euler-Lagrange equations of motion:
\begin{equation}
\ddot{q}_1^\dagger = - \frac{\alpha^2 c^2}{L^2} q_1^\dagger; \qquad \ddot{q}_2 = - \frac{\alpha^2 c^2}{L^2} q_2
\end{equation}
In terms of these two complex variables, the aikyon behaves like two independent complex-valued oscillators. However, the degrees of freedom of the aikyon couple with each other when expressed in terms of the self-adjoint variables $q_B$ and $q_F$. This is because $q_1$ and $q_2$ both depend on $q_B$ and $q_F$, the difference being that $q_1$ depends on $\beta_1$ and $q_2$ depends on $\beta_2$.

The trace Hamiltonian for the aikyon is 
\begin{equation}
Tr{\cal H} = Tr [p_1 \dot{q}_1^\dagger + p_2 \dot{q}_2 - Tr {\cal  L}] = \frac{a_1 a_0}{2}Tr \left[ \frac{4}{a^2_1 a^2_0}p_1 p_2 + \frac{\alpha^2 c^2}{L^2} q_1^\dagger  q_2 \right]
\end{equation}
and Hamilton's equations of motion are \cite{q1q2uni}
\begin{equation}
\dot{q}_1^\dagger  = \frac{2}{a_1 a_0} p_2  \qquad  \dot {q}_2  = \frac{2}{ a_1 a_0 } p_1  ;  \qquad
\dot{p}_1 = -\frac{a_1 a_0 \alpha^2  c^2}{2L^2} q_2; \qquad \dot{p}_2 = -\frac{a_1 a_0 \alpha^2  c^2}{2L^2} q_1^\dagger
\end{equation}
Because the trace Lagrangian is invariant under global unitary transformations, it possesses a novel conserved charge, known as the Adler-Millard charge, and defined as follows:
\begin{equation}
\tilde{C} = \sum_i [q_{Bi}, p_{Bi}] - \sum_i\{q_{Fi,} p_{Fi}\}
\end{equation}
which is the sum, over all bosonic degrees of freedom, of the commutator $[q_B, p_B]$ minus the sum, over all fermionic degrees of freedom, of the 
anti-commutator $\{q_F, p_F\}$. This charge plays an important role in the theory, and is responsible for emergence of quantum field theory at energies below the Planck scale. For this model, we have simply
\begin{equation}
{C}=[q_1, p_1] + [q_2, p_2]
\end{equation}
If we express $q_1^\dagger$ and $q_2$ in terms of their self-adjoint and anti-self-adjoint parts  as $q_1^\dagger = q_1^s + q_1^{as}$, $q_2 = q_2^s + q_2^{as}$, then we can use the self-adjoint and anti-self-adjoint parts as new dynamical variables \cite{q1q2uni},
\begin{equation}
q_1^s= \frac{1}{2}(q_1 + q_1^{\dagger}) ; \qquad q_1^{as} = \frac{1}{2}(q_1^\dagger-q_1) ; \qquad q_2^s= \frac{1}{2}(q_2 + q_2^{\dagger})  ;  \qquad q_2^{as}= \frac{1}{2}(q_2 - q_2^{\dagger})
\end{equation}
and then the Lagrangian  becomes
\begin{equation}
{\cal L} =  \frac{a_1 a_0}{2} Tr \bigg[ \frac{4}{a^2_1 a^2_0} \bigg\{ \dot{q}_2^s \dot{q}_1^s  + \dot{q}_2^{as} \dot{q}_1^{as} \bigg\}  
-\frac{\alpha^2 c^2}{L^2} \bigg\{q_1^s q_2^s + q_1^{as} q_2^{as} \bigg\} +   \frac{4}{a^2_1 a^2_0} \bigg\{ \dot{q}_2^s \dot{q}_1^{as}  + \dot{q}_2^{as} \dot{q}_1^{s} \bigg\}  
-\frac{\alpha^2 c^2}{L^2} \bigg\{q_1^s q_2^{as} + q_1^{as} q_2^{s}  \bigg\} \bigg]
\label{waahwaah}
\end{equation}
In terms of these new variables, and in terms of the self-adjoint and anti-self-adjoint parts of the momenta $p^s$ and $p^{as}$, 
the real and imaginary parts of the trace Hamiltonian are given as follows \cite{q1q2uni}
\begin{equation}
Tr H^s = \frac{a_1 a_0}{2} Tr \bigg[ \frac{4}{a^2_1 a^2_0} \bigg\{ p_1^s p_2^s  + p_1^{as} p_2^{as} \bigg\}  
+\frac{\alpha^2 c^2}{L^2} \bigg\{q_1^s q_2^s + q_1^{as} q_2^{as}  \bigg\} \bigg]
\label{h1}
\end{equation}
\begin{equation}
Tr H^{as} = \frac{a_1 a_0}{2} Tr \bigg[ \frac{4}{a^2_1 a^2_0} \bigg\{ p_1^s p_2^{as}  + p_1^{as} p_2^{s} \bigg\}  
+\frac{\alpha^2 c^2}{L^2} \bigg\{q_1^s q_2^{as} + q_1^{as} q_2^{s}  \bigg\} \bigg]
\label{him}
\end{equation}
The  self-adjoint and anti-self-adjoint components of the Adler-Millard charge become
\begin{equation}
\begin{split}
&\tilde{C}^{as} = [q_1^s, p_1^{s}] + [q_1^{as}, p_{1}^{as}] +  [q_2^s, p_2^{s}] +  [q_2^{as}, p_2^{as}] \\
& \tilde{C}^{s} = [q_1^s, p_1^{as}] + [q_1^{as}, p_{1}^{s}] +  [q_2^s, p_2^{as}] +  [q_2^{as}, p_2^{s}]
\end{split}
\end{equation}
We can now appreciate the significance of the anti-self-adjoint part of the Hamiltonian. It causes non-unitary evolution, equivalent to rapid variations in the Hamiltonian across the matrix dynamics phase space.
The ratio $L_P/L$, which appears in the various terms, plays a very important role. If $L\gg L_P$, the variations are ignorable on scales much larger than Planck length, and hence can be justifiably averaged, so as to arrive at the emergent theory, which is in fact quantum theory. And indeed, on the scales at which current experiments are being conducted, $L$ is much larger than Planck length, and the anti-self-adjoint part of the Hamiltonian can be safely neglected. We are completely justified in using the laws of quantum field theory at these scales.

In the theory of trace dynamics, at energies below Planck scale, quantum theory (without a background spacetime) emerges. This is achieved by coarse-graining the theory over many Planck time scales, and by applying the methods of statistical thermodynamics to arrive at the emergent quantum theory. This happens provided the self-adjoint part of the Adler-Millard charge can be neglected, and the anti-self-adjoint part of the Hamiltonian can be neglected. When that happens, the Adler-Millard charge gets equipartitioned over the four degrees of freedom, the equipartitioned value is identified with Planck's constant, and quantum commutation relations emerge, for the statistically averaged dynamical variables at statistical equilibrium:
\begin{equation}
[q_1^s, p_1^{s}]=i\hbar, \qquad [q_1^{as}, p_{1}^{as}]=i\hbar, \qquad [q_2^s, p_2^{s}] = i\hbar, \qquad [q_2^{as}, p_2^{as}]=i\hbar
\end{equation}
The averaged dynamical variables obey Heisenberg equations of motion. An equivalent Schr\"{o}dinger picture can also be constructed. Working in this framework, we have demonstrated the existence of a ground state \cite{q1q2uni}  in this emergent theory, which we call spontaneous quantum gravity. This ground state possibly has significant implications for the issue of singularity avoidance in quantum cosmology. The analysis reported there could also assist in working out the running of the coupling constant $\alpha$ as a function of the coarse-graining scale.
 
 One can from here make the transition to the $q_B$ and $q_F$ dynamical variables and write the quantum theory in terms of those variables. We point out that if one starts right from the outset with $q_B$ and $q_F$ variables, and does Lagrangian dynamics with them, care has to be taken to assume $\beta_1$ and $\beta_2$ as constant Grassmann c-numbers. And not constant fermionic matrices, as was done by us in \cite{maithresh2019} and \cite{MPSingh}. There, we assumed these matrices to be invariant under general unitary transformations, so as to arrive at the conserved Adler-Millard charge. It seems however, that the resulting charge, despite appearances, does not show up as a conserved quantity, when the solution to the equations of motion is substituted in the expression for the charge [I am grateful to Roy and Sahu for this observation \cite{RS2020}]. And the culprit is precisely the fact that we chose $\beta_1$ and $\beta_2$ as matrices. However, all the earlier results  of \cite{maithresh2019, MPSingh} continue to hold so long as we assume $\beta_1$ and $\beta_2$  to be fermionic c-numbers rather than matrices.

If we keep the total time derivative terms in the Lagrangian, we can analogously write the equations of motion for the three (interacting) fermion generations.  The octonionic quantum mechanics will be described by the exceptional Jordan algebra.

There are circumstances though, when the anti-self-adjoint part of the Hamiltonian becomes important. This happens if the electro-colour interactions cause a sufficiently large number of aikyons to get entangled with each other. The entangled system has an associated effective length $L_{eff} \sim L/N$ where $N$ is the number of entangled aikyons. If $N$ is sufficiently large, the effective length goes below Planck length. As a result, we get that $L_P / L_{eff}$ becomes larger than Planck length. Hence the approximation that the variations in the anti-self-adjoint part can be coarse-grained over to arrive at the emergent quantum theory is no longer valid. Superpositions of quantum states break down on laboratory time-scales. This is the process of spontaneous localisation which causes the classical limit to emerge. We re-emphasise that the anti-self-adjoint part was not added to the Lagrangian by hand in an ad-hoc way. Having it in the Lagrangian is essential in order to have a unified theory of interactions. This gives a fundamental origin for the Ghirardi-Rimini-Weber model of objective collapse. 

 \section{Spontaneous localisation}
 In earlier papers \cite{stcw, maithresh2019, MPSingh} we have sketched preliminary ideas as to how spontaneous localisation of fermions gives rise to the emergence of a classical space-time. Here we recall those ideas, and explain what new things we learn from the present paper, now that we know that the underlying space is octonionic. Let us try to understand which terms in the Lagrangian can contribute to the anti-self-adjoint part of the Lagrangian. We would require such terms to also have a component in the lower half `real' quaternionic plane, for them to have a role in the emergence of 4-D space-time. The first term $T_1$ is self-adjoint; hence it is not expected to directly contribute to spontaneous localisation. This possibly explains why gravitation is not  localised; the weak interaction is short range because the associated bosons are massive. Prior to the electro-weak symmetry breaking the weak interaction too would be long range. The $U(1)$ gauge interaction is associated with a self-adjoint operator [the number operator made from the electro-colour ladder operators]. This possibly explains why the electromagnetic interaction is not localised either. The $SU(3)_c$ gauge bosons span only a part of the lower-half plane, belonging `mostly' to the upper-half plane, having originally been assigned to the purely imaginary directions $(e_3, e_5, e_6, e_7)$. It remains to be investigated if the Bose-Einstein nature of the bosonic statistics is playing a role in bosons not being localised [unless they are massive]. It seems clear though that bosons by themselves cannot cause / undergo spontaneous localisation.
 Similarly, it is not clear to us if the terms $T_4$ and $T_8$ take part in causing spontaneous localisation. Most likely, these terms, being potentially three of the four Higgs bosons - are those three Higgs particles which are consumed to give masses to particles. Hence these terms will not be present in the low energy universe.
 
 The role of causing spontaneous localisation then rests with the terms $T_2, T_3, T_6, T_7$ which describe interactions of the fermions with the bosons. Because of the fermions, these terms span the entire octonionic plane. They all have a self-adjoint part as well as an anti-self-adjoint part. Thus, they contribute to the anti-self-adjoint part of the Hamiltonian, while also having a self-adjoint part. Hence they provide an ideal set-up for collapse to take place when sufficiently many fermions get entangled with each other so as to make `larger than Planck length scale' imaginary variations in the Hamiltonian. This localises the fermions to one or the other specific eigenvalues of the self-adjoint part of the fermionic Grassmann matrices. This is how a set of entangled fermions acquires an emergent classical position, giving rise to an emergent 4-D space-time.  The space-time is 4-D because it is in fact the quaternionic subset of the $F_4$ symmetry of the octonions, and it has the desired local Lorentz symmetry, given by ${\rm Aut}(\{\mathbb H\})$ and the related $C\ell(2)$ which generates $SL(2,C)$. This is the part of the symmetry that is common across the three $G_2$ groups of symmetries of the  three fermion generations. The Lorentz symmetry is already present in the Planck scale theory. It emerges as classical precisely in the same way that classical Maxwell electrodynamics emerges from quantum electrodynamics in the macroscopic limit.
 
 How does gravitation emerge? Local Lorentz symmetry of the emergent space-time is generated by spontaneous localisation of one set of entangled fermions, to one specific eigenvalue of $\dot{q}_F / q_F$. It forces the coupled bosonic field also to a specific eigen-value of $\dot{q}_B$. A different set of localised entangled fermions gives rise to another eigenvalue of the coupled Lorentz field. And so on. That is how the Lorentz symmetry is gauged, giving rise to gravity. However, the localisation is of the squared Dirac operator, $Tr [D_B]^2$, one per aikyon. It is not the localisation of $Tr [D_B]$. Hence we infer a spin 2 gravitational field, not a spin one Lorentz field. We explain in some detail in \cite{maithresh2019} how gravitation emerges, as also the laws of classical general relativity. Though at that time we did not realise that $\dot{q}_B$ describes a spin one massless boson, nor that we need octonions. It is clear that we must not quantise the gravitational field. Gravitation  is an emergent collective phenomenon. We should quantise the spin one Lorentz field; rather, the gravito-weak field. The metric is already there in the underlying Lagrangian, in the term $T_1$. See also the important work of Landi and Rovelli on eigenvalues of the Dirac operator as dynamical observables of general relativity \cite{Landi1999, Rovelli}. See also the very interesting work of Zubkov \cite{Zubkov} on `Gauge theory of Lorentz group as a source of the dynamical electroweak symmetry breaking' which possibly has an intimate connection with our present analysis.
 
   In a similar way, the coupling of fermions to the electro-colour degrees of freedom sends them to specific eigenvalues, and the collective set of eigenvalues describes the gauging of electro-colour interaction and emergence of Yang-Mills fields as internal symmetries.
   
   Where are the other four dimensions? Classical space-time is being kept classical and 4-D by the rapid stochastic variations in the other four octonionic dimensions. The variations are stochastic because they are taking place in the aikyonic Hamiltonian of every one of the aikyons in an entangled set, in an uncorrelated manner. That is how we do not directly see the other four octonion directions. This is also the origin of the stochastic noise in collapse models, whose collapse parameters and the spectrum of the CSL noise should now be predictable from the underlying theory. This picture matches very closely with Adler's suggestion that the origin of the CSL noise is a stochastic imaginary component of the 4-D space-time metric. What we have obtained from the underlying theory is very similar to Adler's proposal \cite{Adler2014}. In fact it is possible that the underlying octonionic space has a Poincar\'e symmetry, of which the 4-D Lorentz symmetry is a part, and these stochastic variations belong to the non-Lorentz part of the Poincar\'e group [translations, spin, torsion, and possibly a connection between strong interactions and torsion]. This also explains why the symmetry group of general relativity is the Lorentz group, and not the Poincar\'e group [only rotations, no translations]. Whereas the symmetry group of relativistic quantum mechanics is the Poincar\'e group, not the Lorentz group. Because quantum systems live in eight dimensions, not four. The other four directions, and the associated translation part of the Poincar\'e symmetry, are essential for a meaningful description of spin \cite{Singhspin}. 
   
   A quantum system which has not undergone collapse necessarily lives in 8-D non-commutative octonionic space, {\it not} in 4-D space-time. Although for most purposes the description on a 4-D spacetime background suffices and agrees with experiments done to date. However we have made predictions [e.g. the Lorentz boson, the gravito-weak unification] which are testable, and which will give evidence for the 8-D space in which quantum systems reside. Moreover, we know that the 4-D description leads to puzzles at times - quantum non-locality, and the mysterious nature of spin. These puzzles go away in the 8-D non-commutative description. Also, there is no need for a Kaluza-Klein style spontaneous compactification of the extra four dimensions. For classical systems, spontaneous localisation effectively suppresses the extra four dimensions. Whereas quantum systems actually probe the other four dimensions - they must not be compactified. The symmetry group of the unified theory is $F_4$, the group of automorphisms of the 8-D octonionic space, for three generations of fermions.
   
     Once a classical space-time background is available, the emergent quantum theory can be related to conventional quantum field theory. We believe that under this situation our emergent theory [evolution in Connes time] coincides with the Horwitz-Stueckelberg covariant formulation of relativistic quantum mechanics \cite{stueck}. When they treat time part of spacetime also on the same footing as spacetime, and hence in a fully covariant manner, they need to introduce an extrinsic time parameter. It is plausible that this parameter is identical with Connes time.
 
 \section{Outlook}
 \subsection {Testable predictions}
 Our theory has no free parameters and is falsifiable. If the predicted Lorentz boson is ruled out by experiments, our theory will be ruled out. Also, our theory predicts that there are no new particles to be discovered, apart from the Lorentz boson. Thus the discovery of other new particles, say a fourth generation of fermions, will also rule out our theory. Elsewhere, we have predicted the experimentally testable Karolyhazy length uncertainty relation as a consequence of our theory. This says that if a device is used to measure a length $L$, there will be a minimum uncertainty $\Delta L$, given by the Karolyhazy relation \cite{Karolyhazy:74, Karolyhazy:1982, Karolyhazy:90, Karolyhazy:95, Ng, Amelio:1999, Singh:KL} 
 \begin{equation}
 (\Delta L)^3 \sim L_P^2 \; L
 \end{equation}
 A dedicated experiment is planned to test this relation \cite{s2020experiment}. We also note that this relation implies holography: quantum information in a region of size $L$ grows as area of the region's boundary, not as the region's volume. This is because if $\Delta L$ is the smallest possible linear extent of a cell with one unit of information, then it follows from this relation that $L^3 / (\Delta L)^3 \sim L^2 / L_P^2$ which of course is holography. Note that the minimum of length in our theory is not $L_P$, but the much larger $\Delta L$ given by the above relation. A holographic theory of quantum gravity must necessarily predict and satisfy this holographic relation. 
 
 We have also predicted the unification of gravity and the weak interaction. This implies a variation in the value of $G_N$ which must be looked for at small scales \cite {Onofrio1, Onofrio2}. Our theory also predicts the phenomenon of spontaneous localisation, currently being tested in the laboratory \cite{RMP:2012, Mauro2019}.  
   We predict a specific model of Continuous Spontaneous Localisation [CSL] in the class of objective collapse models. The collapse rate $\lambda$ is given by $(L_P/L)^3 \; \tau_P^{-1}$ where $L$ is the Compton wavelength of the proton. Numerically, this is of the order of $10^{-17}$ s$^{-1}$ which agrees with the value proposed by the Ghirardi-Rimini-Weber-Pearle theory. The collapse rate in our theory grows as cube of the mass in the system; hence this is not the mass-proportional CSL model. We are currently investigating the CSL noise spectrum predicted by the underlying aikyon theory - this will aid comparison of the aikyon theory versus experiment.

  In addition, we have predicted the testable novel phenomenon of spontaneous localisation in time \cite{Singh:2019}. Thus, our theory makes several predictions, testable with current technology, which can be used to confirm or rule out the theory. We have also explained the remarkable fact the the Kerr-Newman black hole has the same gyromagnetic ratio as the electron \cite{MPSingh}.
 
 It would also be interesting to explore if there is a left-over Lorentz radiation background [made of Lorentz bosons]  from the beginnings of the universe, analogous to the cosmic microwave background. And whether this radiation could have some role to play as dark energy? We have in fact recently suggested, based on our theory, that dark energy is a large scale quantum gravitational phenomenon \cite{Singh:DE}.
 
 \subsection{Concluding Remarks}
 
 The implications of the existence of Connes time remain to be understood. The most likely point of  contact with conventional quantum field theory is the Horwitz-Stueckleberg fully covariant formulation of relativistic quantum mechanics, which introduces an additional external time parameter. It is entirely possible that this additional parameter is the same as Connes time \cite{stueck}. 
 
 The Higgs mechanism and the running of the coupling $\alpha$ as well as values of standard model parameters remain to be worked out.
 
 The underlying Planck scale matrix dynamics is deterministic \cite{Singhdice}. It does not violate Bell's inequalities, because of the non-commutative nature of the octonionic space. There is Lorentz invariance but the space is non-commutative. Hence one should not invoke emergent concepts such as locality and non-locality. There is an EPR influence and correlation, but it is not `outside' the light-cone, because there is no light-cone to begin with. Also, it is not a hidden variables theory. The deterministic evolution is in general non-unitary, and when non-unitarity is significant it will lead to breakdown of superpositions. Quantum indeterminism arises in the coarse-gained emergent theory precisely because of the coarse-graining. We do not have information on regions that have been coarse-grained over. This effectively converts the underlying deterministic theory into an apparently non-deterministic emergent theory. The Born probability rule arises because the norm of the state-vector is preserved in the underlying trace dynamics evolution, in spite of it being non-unitary evolution. This happens because the evolution of an aikyon in Connes time, in the non-commutative octonionic space, is `geodesic / free-fall'. It is a closed system.
 
 We believe we have proposed a highly promising and falsifiable unified theory of interactions. Interactions are always mediated by spin one bosons, including the Lorentz boson. The Lorentz symmetry, and not gravity, is unified with the standard model interactions.  We sincerely hope other researchers will also investigate this approach and take it to its logical conclusion.
 
 The idea that quantum theory and gravitation both are emergent phenomena is not new. It is being seriously pursued by other researchers as well. See in particular the work of Vitaly Vanchurin \cite{Vitaly} and also of Ricardo Torrome \cite{Torrome}. Also, Stephen Adler concludes his book \cite{Adler:04} with the following remarks:
 
 ``To conclude, we believe that the kinematical framework of trace dynamics, as developed in this book, provides a fruitful new direction for exploration in the search for a unifying theory of particles and forces. Specifically, if quantum theory is an emergent phenomenon, then the sought for unification of gravitation with the quantum field theoretic standard model may not involve ``quantising" gravitation. Rather, the ideas developed in this book suggest, one should seek a common origin for both gravitation and quantum field theory at the deeper level of physical phenomena from which quantum field theory emerges".
 
 After this paper was written it was brought to my attention by Thanu Padmanabhan that the concept of an atom of space-time-matter / aikyon is present also in the very comprehensive works of Kazunari Shima. See Ref. \cite{Shima, Shima2} and references therein. Shima calls his theory Supersymmetric Non-linear General Relativity. The goal is essentially the same as ours, a first principles unification of space-time geometry and the standard model. We intend to explore possible connections between these two approaches. An octonionic unification of gravitation and the standard model has been proposed also by Perelman \cite{Perelman}, again with strong overlaps with our research. See also, `The exceptional Jordan algebra and the matrix string' by Smolin \cite{Smolin}. 
 
 \bigskip
 \noindent{\bf Acknowledgements:} The author would like to thank Abhinash Kumar Roy and Anmol Sahu for collaboration, and for intense and helpful discussions. The author would also like to thank Stephen Adler, Cohl Furey, Niels Gresnigt, and Ovidiu Cristinel Stoica, for helpful correspondence, and support and encouragement during the course of this project. Thanks also to Basudeb Dasgupta, Debajyoti Choudhury, Roberto Onofrio, Thanu Padmanabhan, Roberto Percacci and Carlos Perelman for helpful comments on an earlier version of the manuscript, and for drawing my attention to related relevant research.
 \bigskip
\vskip 0.4 in

\noindent {\bf Note added in proofs}: A recent discussion \cite{pcd} motivates us to drop the adjointness requirement on $\dot{q}_B$ in (40) and (41) [except the $e_0$ component, which is still assumed to be real], as a result of which there are now two Lorentz bosons in the theory. There are then 32 bosonic degrees of freedom in the theory, which matches the 32 fermionic degrees of freedom per generation. The four additional degrees of freedom coming from the two Lorentz bosons raise the standard model bosonic degrees of freedom from 28 to 32. This (32, 32) match between bosons and fermions, which results {\it only after} unifying  gravity with the standard model, is highly encouraging. The author thanks Debajyoti Choudhury and Basudeb Dasgupta for discussions which motivated this conclusion. Eqn. (41) has terms which suggest the Lorentz interaction could violate parity. This is under investigation. 

\bigskip

\centerline{\bf REFERENCES}
\bibliographystyle{unsrt}
\bibliography{biblioqmtstorsion}

\def\polhk#1{\setbox0=\hbox{#1}{\ooalign{\hidewidth
  \lower1.5ex\hbox{`}\hidewidth\crcr\unhbox0}}} \def\cprime{$'$}
  \def\cprime{$'$}
\begin{thebibliography}{10}

\bibitem{Singh:2012}
T.~P. Singh.
\newblock The problem of time and the problem of quantum measurement.
\newblock In T.~Filk and A.~von Muller, editors, {\em Re-thinking time at the
  interface of physics and philosophy}, (arXiv:1210.81110).
  Berlin-Heidelberg:Springer, 2015.

\bibitem{Connes2000}
Alain Connes.
\newblock {\em Visions in Mathematics - GAFA 2000 Special volume, Part II},
  chapter Non-commutative geometry 2000, pages 481 Eds. N. Alon, J. Bourgain,
  A. Connes, M. Gromov and V. Milman, arXiv:math/0011193.
\newblock Springer, 2000.

\bibitem{Adler:04}
Stephen~L. Adler.
\newblock {\em Quantum theory as an emergent phenomenon}.
\newblock Cambridge University Press, Cambridge, 2004.

\bibitem{Adler:94}
Stephen~L. Adler.
\newblock Generalized quantum dynamics.
\newblock {\em Nucl. Phys. B}, 415:195, 1994.

\bibitem{AdlerMillard:1996}
Stephen~L. Adler and Andrew~C. Millard.
\newblock Generalised quantum dynamics as pre-quantum mechanics.
\newblock {\em Nucl. Phys. B}, 473:199, 1996.

\bibitem{maithresh2019b}
Maithresh Palemkota and Tejinder~P. Singh.
\newblock Black hole entropy from trace dynamics and non-commutative geometry.
\newblock arXiv:1909.02434v2 [gr-qc], 2019 submitted for publication.

\bibitem{RMP:2012}
Angelo Bassi, Kinjalk Lochan, Seema Satin, Tejinder~P. Singh, and Hendrik
  Ulbricht.
\newblock Models of wave function collapse, underlying theories, and
  experimental tests.
\newblock {\em Rev. Mod. Phys.}, 85:471 arXiv:1204.4325 [quant--ph], 2013.

\bibitem{connes1994}
Alain Connes and Carlo Rovelli.
\newblock von {N}eumann algebra automorphisms and time-thermodynamics relation
  in general covariant quantum theories.
\newblock {\em Class. Quant. Grav.}, 11:2899, 1994.

\bibitem{Takesaki1}
Masamichi Takesaki.
\newblock Theory of operator algebras {II}.
\newblock In {\em Encylopedia of Mathematical Sciences}, volume 125. Springer
  Verlag, Berlin, 2003.

\bibitem{Takesaki2}
Masamichi Takesaki.
\newblock Tomita's theory of modern {H}ilbert algebras and its applications.
\newblock In {\em Lecture Notes in Mathematics}, volume 128. Springer, Berlin,
  1970.

\bibitem{Nykodym}
O.~Nykodym.
\newblock Sur une g{\'e}n{\'e}ralisation des int{\'e}grales de {M}. {J}.
  {R}adone.
\newblock {\em Fundamenta Mathematicae}, 15:131--179, 1930.

\bibitem{Singh:sqg}
Tejinder~P. Singh.
\newblock Spontaneous quantum gravity.
\newblock arXiv:1912.03266v2, 2019 [submitted for publication].

\bibitem{singh2019qf}
Tejinder~P. Singh.
\newblock From quantum foundations to spontaneous quantum gravity: an overview
  of the new theory.
\newblock {\em Zeitschrift f\"ur Naturforschung A}, arXiv:1909.06340
  [gr-qc]:DOI: https://doi.org/10.1515/zna--2020--0073, 2020.

\bibitem{maithresh2019}
Maithresh Palemkota and Tejinder~P. Singh.
\newblock Proposal for a new quantum theory of gravity {III}: Equations for
  quantum gravity, and the origin of spontaneous localisation.
\newblock {\em Zeitschrift f\"ur Naturforschung A}, 75:143, 2019
  DOI:10.1515/zna-2019-0267 arXiv:1908.04309.

\bibitem{Singhspin}
Tejinder~P. Singh.
\newblock Octonions, trace dynamics and non-commutative geometry: a case for
  unification in spontaneous quantum gravity.
\newblock {\em Zeitschrift f\"ur Naturforschung A},
  arXiv:2006.16274v2:https://doi.org/10.1515/zna--2020--0196, 2020.

\bibitem{Chams:1997}
Ali~H. Chamseddine and Alain Connes.
\newblock The spectral action principle.
\newblock {\em Commun. Math. Phys.}, 186:731 arXiv:hep--th/9606001, 1997.

\bibitem{MPSingh}
Meghraj~M S, Abhishek Pandey, and Tejinder~P. Singh.
\newblock Why does the {Kerr-Newman} black hole have the same gyromagnetic
  ratio as the electron?
\newblock {\em submitted for publication}, arXiv:2006.05392, 2020.

\bibitem{jacobson_1995}
Ted Jacobson.
\newblock Thermodynamics of spacetime: The {E}instein equation of state.
\newblock {\em Physical Review Letters}, 75(7):1260--1263 arXiv:grqc/9505004,
  Aug 1995.

\bibitem{paddy}
T.~Padmanabhan.
\newblock Gravity and is thermodynamics.
\newblock {\em Current Science}, 109:2236, arXiv:1512.06546, 2015.

\bibitem{Schucker2000spin}
Thomas Schucker.
\newblock Spin group and almost commutative geometry.
\newblock hep-th/0007047, 2000.

\bibitem{Singh:2019}
Tejinder~P. Singh.
\newblock Space-time from collapse of the wave-function.
\newblock {\em Zeitschrift f\"ur Naturforschung A}, 74:147
  arXiv.org:1809.03441, 2019.

\bibitem{Dixon}
Geoffrey~M. Dixon.
\newblock {\em Division algebras, octonions, quaternions, complex numbers and
  the algebraic design of physics}.
\newblock Kluwer, Dordrecht, 1994.

\bibitem{Gursey}
C.~H. Tze and F.~Gursey.
\newblock {\em On the role of division, {Jordan} and related algebras in
  particle physics}.
\newblock World Scientific Publishing, 1996.

\bibitem{f1}
Cohl Furey.
\newblock Standard model physics from an algebra? {Ph. D.} thesis, university
  of {Waterloo}.
\newblock arXiv:1611.09182 [hep-th], 2015.

\bibitem{f2}
Cohl Furey.
\newblock Three generations, two unbroken gauge symmetries, and one
  eight-dimensional algebra.
\newblock {\em Phys. Lett. B}, 785:1984, 2018.

\bibitem{f3}
Cohl Furey.
\newblock ${SU(3)_C\times SU(2)_L \times U(1)_Y (\times U(1)_X)}$ as a symmetry
  of division algebraic ladder operators.
\newblock {\em Euro. Phys. J. C}, 78:375, 2018.

\bibitem{Chisholm}
J.~Chisholm and R.~Farwell.
\newblock Clifford geometric algebras: with applications to physics,
  mathematics and engineering.
\newblock page 365. Birkhauser, Boston, 1996 Ed. W. R. Baylis.

\bibitem{Trayling}
G.~Trayling and W.~Baylis.
\newblock A geometric basis for the standard-model gauge group.
\newblock {\em J. Phys. A: Math. Theor.}, 34:3309, 2001.

\bibitem{Dubois_Violette_2016}
Michel Dubois-Violette.
\newblock Exceptional quantum geometry and particle physics.
\newblock {\em Nuclear Physics B}, 912:426--449, Nov 2016.

\bibitem{Todorov:2019hlc}
Ivan Todorov.
\newblock {Exceptional quantum algebra for the standard model of particle
  physics}.
\newblock {\em Nucl. Phys. B}, 938:751 arXiv:1808.08110 [hep--th], 11 2019.

\bibitem{Dubois-Violette:2018wgs}
Michel Dubois-Violette and Ivan Todorov.
\newblock {Exceptional quantum geometry and particle physics II}.
\newblock {\em Nucl. Phys. B}, 938:751--761 arXiv:1808.08110 [hep--th], 2019.

\bibitem{Todorov:2018yvi}
Ivan Todorov and Svetla Drenska.
\newblock {Octonions, exceptional Jordan algebra and the role of the group
  $F_4$ in particle physics}.
\newblock {\em Adv. Appl. Clifford Algebras}, 28(4):82 arXiv:1911.13124
  [hep--th], 2018.

\bibitem{Todorov:2020zae}
Ivan Todorov.
\newblock {J}ordan algebra approach to finite quantum geometry.
\newblock In {\em PoS}, volume CORFU2019, page 163, 2020 DOI:
  10.22323/1.376.0163.

\bibitem{ablamoowicz}
Rafal Ablamowicz.
\newblock Construction of spinors via {W}itt decomposition and primitive
  idempotents: A review.
\newblock In Rafal Ablamowicz and P.~Lounesto, editors, {\em Clifford algebras
  and spinor structures}, page 113. Kluwer Acad. Publ., 1995.

\bibitem{baez2001octonions}
John~C. Baez.
\newblock The octonions.
\newblock {\em arXiv:math/0105155}, 2001.

\bibitem{Baez_2011}
John~C. Baez.
\newblock Division algebras and quantum theory.
\newblock {\em Foundations of Physics}, 42(7):819--855, May 2011.

\bibitem{baez2009algebra}
John~C. Baez and John Huerta.
\newblock The algebra of grand unified theories, 2009 arXiv:0904.1556 [hep-th].

\bibitem{Baez}
John~C. Baez and John Huerta.
\newblock Division algebras and supersymmetry {II}.
\newblock {\em Adv. Math. Theor. Phys.}, 15:1373, 2011.

\bibitem{Jordan}
P.~Jordan, John von Neumann, and E.~Wigner.
\newblock On an algebraic generalisation of the quantum mechanical formalism.
\newblock {\em Ann. Math.}, 35:29, 1934.

\bibitem{Albert1933}
A.~Adrien Albert.
\newblock On a certain algebra of quantum mechanics.
\newblock {\em Annals of Mathematics}, 35:65, 1933.

\bibitem{Gunaydin2}
M.~Gunaydin and F.~Gursey.
\newblock Quark structure and octonions.
\newblock {\em J. Math. Phys.}, 14:1651, 1973.

\bibitem{Stoica}
Ovidiu~Cristinel Stoica.
\newblock The standard model algebra ({Leptons}, quarks and gauge from the
  complex algebra {Cl(6)}).
\newblock {\em Advances in Applied Clifford Algebras}, arXiv:1702.04336(28):52,
  2018.

\bibitem{Gillard_2019}
Adam~B. Gillard and Niels~G. Gresnigt.
\newblock Three fermion generations with two unbroken gauge symmetries from the
  complex sedenions.
\newblock {\em The European Physical Journal C}, 79(5), May 2019
  arXiv:1904.03186.

\bibitem{Yokota}
Ichiro Yokota.
\newblock Exceptional {L}ie groups.
\newblock arXiv:0902.043 [math.DG], 2009.

\bibitem{tkey}
Ivan Todorov and Michel Dubois-Violette.
\newblock Deducing the symmetry of the standard model fom the automorphism and
  structure groups of the exceptional {J}ordan algebra.
\newblock arXiv:1806.9450 [hep-th], 2018.

\bibitem{q1q2uni}
Abhinash~Kumar Roy, Anmol Sahu, and Tejinder~P. Singh.
\newblock Trace dynamics, and a ground state in spontaneous quantum gravity.
\newblock www.tifr.res.in/\string~tpsingh/q1q2uni.pdf:Submitted for
  publication, 2020 [available at home page of TPS].

\bibitem{Ilka}
Ilka Agricola.
\newblock Old and new in the exceptional group $\{G_2\}$.
\newblock {\em Notices of the AMS}, 55:922, 2008.

\bibitem{Onofrio1}
Roberto Onofrio.
\newblock On weak interactions as short distance manifestations of gravity.
\newblock {\em Mod. Phy. Letts. A}, 28:1350022 arXiv:1412.4513 [hep--ph], 2013.

\bibitem{Onofrio2}
Roberto Onofrio.
\newblock Proton radius puzzle and quantum gravity at the {F}ermi scale.
\newblock {\em Europhysics Letters}, 104:20002 arXiv:1312.3469 [hep--ph], 2013.

\bibitem{Percacci}
Fabrizio Nesti and Roberto Percacci.
\newblock Gravi-weak unification.
\newblock {\em J. Phys. A}, 41:075405 arXiv:0706.3307, 2008.

\bibitem{Percacci2}
Kirill Krasnov and Roberto Percacci.
\newblock Gravity and unification: a review.
\newblock {\em Class. Quant. Grav.}, 35:143001 arXiv:1712.03006 [hep--th],
  2018.

\bibitem{Singhspin1}
Tejinder~P. Singh.
\newblock A basic definition of spin in the new matrix dynamics.
\newblock {\em Zeitschrift f\"ur Naturforschung A}, arXiv:2006.16274v1:DOI:
  https://doi.org/10.1515/zna--2020--0183, 2020.

\bibitem{Cahill:2020lry}
Kevin Cahill.
\newblock {Is the local Lorentz invariance of general relativity implemented by
  gauge bosons that have their own Yang-Mills-like action?}
\newblock {\em Phys. Rev. D}, To appear:arXiv:2008.10381 [gr--qc], 8 2020.

\bibitem{BdS}
A.~Borel and J.~de~Siebenthal.
\newblock Le sou groupes fermes de rang maximum des groupes de lie clos.
\newblock {\em Commentarii Mathematici Helvetici}, 23:200, 1949.

\bibitem{RS2020}
Abhinah~Kumar Roy and Anmol Sahu.
\newblock ({\it {p}rivate {c}ommunication}).
\newblock 2020.

\bibitem{stcw}
Tejinder~P. Singh.
\newblock Space-time from collapse of the wave-function.
\newblock {\em Zeitschrift f\"ur Naturforschung A}, 74:147, arXiv:1809.03441,
  2019.

\bibitem{Landi1999}
Giovanni Landi.
\newblock Eigenvalues as dynamical variables.
\newblock {\em Lect. Notes Phys.}, 596:299, 2002 gr-qc/9906044.

\bibitem{Rovelli}
Giovanni Landi and Carlo Rovelli.
\newblock General relativity in terms of {Dirac} eigenvalues.
\newblock {\em Phys. Rev. Lett.}, 78:3051 arXiv:gr--qc/9612034, 1997.

\bibitem{Zubkov}
M~A Zubkov.
\newblock Gauge theory of {L}orentz group as a source of the dynamical
  electroweak symmetry breaking.
\newblock {\em JHEP}, 1309:044 arXiv:1301.6971, 2013.

\bibitem{Adler2014}
S.~L. Adler.
\newblock Gravitation and the noise needed in objective reduction models.
\newblock arXiv:1401.0353 [gr-qc] 2014.

\bibitem{stueck}
Lawrence~P. Horwitz.
\newblock {\em Relativistic quantum mechanics}.
\newblock Springer Netherlands, 2015.

\bibitem{Karolyhazy:74}
F.~Karolyhazy.
\newblock {\em Magy. Fiz. Foly.}, 22:23, 1974.

\bibitem{Karolyhazy:1982}
F.~Karolyhazy, A.~Frenkel, and B.~Lukacs.
\newblock In A.~Shimony and H.~Feshbach, editors, {\em Physics as natural
  philosophy}. MIT Press, Cambridge, 1982.

\bibitem{Karolyhazy:90}
F.~Karolyhazy.
\newblock In A.~Miller, editor, {\em Sixty-two years of uncertainty}. Plenum,
  New York, 1990.

\bibitem{Karolyhazy:95}
F.~Karolyhazy.
\newblock In M.~Ferrero and A.~van~der Merwe, editors, {\em Fundamental
  problems of quantum physics}. Kluwer Acad. Publ., Netherlands, 1995.

\bibitem{Ng}
Y.~Jack Ng.
\newblock Entropy and gravitation: From black hole computers to dark energy and
  dark matter.
\newblock {\em Entropy}, 21:1035, 2019.

\bibitem{Amelio:1999}
Giovanni Amelino-Camelia.
\newblock Gravity-wave interferometers as quantum gravity detectors.
\newblock {\em Nature}, 398:216, 1999.

\bibitem{Singh:KL}
Tejinder~P. Singh.
\newblock Proposal for a new quantum theory of gravity {V}: Karolyhazy
  uncertainty relation, {P}lanck scale foam, and holography.
\newblock {\em Pramana - Journal of Physics [to appear]}, arXiv:1910.06350,
  2020.

\bibitem{s2020experiment}
Sander Vermeulen, Lorenzo Aiello, Aldo Ejlli, William Griffiths, Alasdair
  James, Katherine Dooley, and Hartmut Grote.
\newblock An experiment for observing quantum gravity phenomena using twin
  table-top 3d interferometers.
\newblock {\em arXiv:2008.04957}, 2020.

\bibitem{Mauro2019}
Matteo Carlesso and Mauro Paternostro.
\newblock Opto-mechanical tests of collapse models.
\newblock arXiv:1906.11041, 2019.

\bibitem{Singh:DE}
Tejinder~P. Singh.
\newblock Dark energy as a large scale quantum gravitational phenomenon.
\newblock {\em Mod. Phy. Letts. A}, 35:2050195 arXiv:1911.02955
  DOI:https://doi.org/10.1142/S0217732320501953, 2020.

\bibitem{Singhdice}
Tejinder~P. Singh.
\newblock Nature does not play dice on the {Planck} scale.
\newblock {\em International Journal of Modern Physics},
  29:https://doi.org/10.1142/S0218271820430129 arXiv:2005.06427, 2020.

\bibitem{Vitaly}
Vitaly Vanchurin.
\newblock The world as a neural network.
\newblock arXiv:2008.01540, 2020.

\bibitem{Torrome}
Ricardo~Gallego Torrome.
\newblock On the origin of the weak equivalence principle in a theory of
  emergent quantum mechanics.
\newblock arXiv:2005.12903, 2020.

\bibitem{Shima}
Kazunari Shima.
\newblock Nonlinear {SUSY} general relativity and significances.
\newblock arXiv:1112.3098 [hep-th]:DOI: 10.1088/1742--6596/343/1/012111, 2011.

\bibitem{Shima2}
Kazunari Shima.
\newblock New {E}instein-{H}ilbert type action of space-time and matter
  -nonlinear-supersymmetric general relativity theory-.
\newblock arXiv:2009.06266 [hep-th], 2020.

\bibitem{Perelman}
Carlos~Castro Perelman.
\newblock ${R\times C\times H\times O}$ valued gravity as a grand unified field
  theory.
\newblock {\em Advances in Applied Clifford Algebras}, 29:22, 2019.

\bibitem{Smolin}
Lee Smolin.
\newblock The exceptional {J}ordan algebra and the matrix string.
\newblock arXiv:hep-th/0104050, 2001.

\bibitem{pcd}
Debajyoti Choudhury and Basudeb Dasgupta.
\newblock [Private Communication, 2020].

\end{thebibliography}

\end{document}